 \documentclass[
twocolumn,showpacs, amsmath, amssymb, floatfix,
nofootinbib,wrapfloat byrevtex]{revtex4}

 \usepackage{amsmath,mathrsfs,graphicx,epstopdf,placeins,float}
 \usepackage{booktabs}
 \usepackage{multirow}
 \usepackage{pstricks}

 \newcommand{\beq}[1]{\begin{equation}\label{#1}}
 \newcommand{\eeq}{\end{equation}}
 \newcommand{\bea}[1]{\begin{eqnarray}\label{#1}}
 \newcommand{\eea}{\end{eqnarray}}

\begin{document} 

 \title{Phase Structure and QNMs of A Charged AdS Dilaton Black Hole}
 \author{Ai-chen Li}
 \email{lac@emails.bjut.edu.cn}
 \author{Han-qing Shi}
\email{hq-shi@emails.bjut.edu.cn}
 \author{Ding-fang Zeng}
 \email{dfzeng@bjut.edu.cn}
 \affiliation{Theoretical Physics Division, College of Applied Sciences, Beijing University of Technology,
 Beijing, China, 100124}
 \begin{abstract}
 We investigate the phase structure of a charged AdS dilaton black hole in the extended phase space which takes the cosmological constant, i.e. the AdS-$\Lambda$ parameter as pressures. Through both thermal ensemble and quasinormal mode analysis, we find that stable phase of the black hole with non-trivial dilaton profiles always exists for both large and small couplings when the AdS-$\Lambda$ is considered dynamical degrees of freedom. This forms a somewhat contrast with previous works which does not do so. Our results provide new examples for the parallelism or equivalences between thermal ensemble methods and dynamic perturbation analysis for black hole phase structures.
 \end{abstract}
\pacs{04.70.Dy, 11.25.Tq, 04.50.Gh, 04.70.Bw}
\maketitle

\section{Introduction}

Anti de-Sitter space/conformal field theory correspondence \cite{Maldacena:1997re}, i.e. AdS/CFT, is a powerful method to unify quantum fields and  gravitations. This correspondence suggests that physics of the gauge field defined on the AdS-boundary can be derived from gravitations in the bulk space and vice versa. Studies on thermodynamics of the AdS black hole are of great interests due to their relevance to thermal field theories defined on the boundary spacetime \cite{Witten:1998qj}. Phase transitions always occur in ordinary thermodynamic systems, so we naturally expect that such phenomenas also appear in black holes according to the AdS/CFT correspondence.  This was firstly discussed by Hawking and Page in \cite{Hawking:1982dh}, where a thermal gas/large AdS black hole phase transition was discovered in simple Einstein gravitations with AdS asymptotics. After that work, similar phase transitions were explored in more gravitation theories with AdS asymptotics \cite{Cai:1998ji, Cvetic:1999ne, Cvetic:2001bk, Cai:2001dz}.
  
Dilaton gravitation theories follow from the low-energy limit of string theory \cite{Gibbons:1987ps}. But in these low energy theories, properties of the black hole change drastically due to the dilaton field. Ref. \cite{Garfinkle:1990qj} gives asymptotically flat dilaton black hole solutions in which the dilaton field is coupled to the Maxwell field in an exponential way. Inspired by AdS/CFT, asymptotically AdS dilaton black holes attract more interests in recent works.  Besides the pure cosmological constant potential, Liouville-type dilaton potential also enters sights of the researchers . This potential arises from the supersymmetry breaking of a higher-dimensional supergravity model with cosmological constant \cite{Gregory:1992kr, Horne:1992bi, Poletti:1994ff}. It is proved that in one and two Liouville type potential models \cite{Mignemi:1991wa,Poletti:1994ff}, neither asymptotically flat nor asymptotical AdS black hole solution is allowed. However, in three potential case, static and charged asymptotically AdS black hole/brane solutions are constructed in \cite{Gao:2004tu, Gao:2004tv, Gao:2005xv}, with some phenomenological applications discussed in \cite{Gao:2006fu} and \cite{Zhang:2015dia}. Ref. \cite{Sheykhi:2009pf} analyses thermodynamic stability and phase structures of such black holes through examinations of the second order thermal quantity $(\partial ^2 M / \partial S^2) _Q$ and shows that in small dilaton coupling case, the system allows no Hawking-Page transition; while in the large  $\alpha$ case, some phases of the system are unstable. Ref. \cite{Sheykhi:2016syb} explores phase transitions in this system by the method of thermodynamic geometries. In this method, divergences of the Ricci scalar coincide with the phase transition points of heat capacities. According to \cite{Sheykhi:2016syb}, the charged AdS dilaton black holes of \cite{Gao:2004tu} enjoy many phase transitions, among which one occurs between two stable black holes. 

Comparing with ordinary thermodynamic systems, the black hole system contains no obvious notion of pressures. However, in ref. \cite{adsLambdaThermo1}\cite{Kubiznak1205}, new perspectives are proposed that pressures can be identified with the negative cosmological constant $P=-\Lambda/8\pi$ $=(d-1)(d-2)/(16\pi L^2)$ with the conjugate volume defined as $V=(\frac{\partial M}{\partial P})_{Q,S}$, and the black hole mass interpreted as enthalpies of the system. In this extended phase space, many new phenomena \cite{ccly1306, panahi1506, panahi1510, panahi1511, Kubiznak1608, panahi1608, fernado1611, panahi1702} related with the AdS charged black holes are explored, e.g. the small/large hole transition  is investigated and interpreted as the van der Waals liquid/gas transition in ordinary thermal systems; other transitions like the solid/liquid/gas ones were also found in Kerr-AdS black holes \cite{Altamirano:2013uqa}; while the multiply reentrant phase transition and triple points are observed in the charged and rotating Born-Infeld black hole systems \cite{Gunasekaran:2012dq}.  To treat the cosmological constant as an independent thermodynamic variable has two reasons. Firstly, just as pointed out by \cite{adsLambdaThermo1, Kubiznak1205} and verified in \cite{ccly1306, panahi1506, panahi1510, panahi1511, Kubiznak1608, panahi1608, fernado1611, panahi1702, Altamirano:2013uqa,Gunasekaran:2012dq}, such a doing uncovers us many new interesting feature of black holes being considered. The second is, the cosmological constant in the real universe is nonzero, although positively valued and many orders of magnitude smaller than expectations. Considering our poor understandings on its origin, assuming independence thermodynamically in general gauge/gravity duality studies is a wisdom and may shed light on this question's resolving.

 Our purpose in this paper is to investigate the phase structure and stability issues involved in the charged dilaton black hole of \cite{Gao:2004tv, Gao:2005xv} in the extended phase space formulation.  According to ref.\cite{Sheykhi:2009pf,Sheykhi:2016syb} the charged dilaton black hole has rich structure of phases, even when the pressure effects are not considered. We expect that more interesting phase transitions will occur in this system when pressures are considered, just as pointed previously. One statement of ref. \cite{Sheykhi:2009pf} is that, the charged dilaton black holes are thermodynamically unstable in the large coupling constant case. It is very natural to ask if such instabilities would still be the case when the cosmological constant is considered a dynamic degree of freedom. We will try to understand this question from both thermo-ensemble and dynamical perturbation analysis. While previous works \cite{Sheykhi:2009pf,Sheykhi:2016syb, zzmz1305, zzmz1403, panahi1503, panahi1509, mlx1601, panahi1609, panahi1703} are mainly from thermo-ensemble aspects. Our dynamic consideration borrow ideas from  ref.\cite{Liu2014gvf,mahap1602,chab1606}, which calculated the quasinormal modes (QNMs hereafter) in the AdS- or asymptotically AdS-RN black holes and uncover interesting links between the thermal phase and quasinormal frequencies (QNFs hereafter). We will see that in the charged AdS dilaton black holes, similar link or parallelisms will also appear, of course somewhat modulated by the dilaton coupling constant.

Our work is structured as follows. This section is the background and motivation introduction. The next section will briefly review the (n+1)-dimensional asymptotically AdS black hole solution of Einstein-Maxwell-dilaton theory with Liouville-type potentials. Section \ref{calThermodynamic} gives definition for all relevant thermodynamic quantities in latter discussions. The generalized first law of black hole thermodynamics in the extend phase space is also checked in this section. The phase structure and the thermal stability issue of the system will be investigated in section \ref{BHThermodynamic}. Section \ref{DynaAna} will be devoted to the QNMs' calculation and possible parallelism/equivalence analysis between the thermal ensemble method and dynamic consideration. The last section is our conclusion and discussions. 

\section{Charged Dilaton Black Holes in AdS Space}
\label{fieldeq}

This section is a review on the asymptotically AdS black hole solution in dilaton gravitation theory and related thermodynamic definition. We begin with the action of (n+1)-dimensional ($n \ge 3$) Einstein-Maxwell-dilaton gravity
\bea{EMD}
&&\hspace{-5mm}S=\frac{1}{16\pi G}\!\!\int\!\!d^{n\scriptscriptstyle+1}\!x\sqrt{-g}[R-\frac{4}{n-1}(\nabla \Phi)^2
\\
&&\rule{15mm}{0pt}-V(\Phi)-e^{-4\alpha \Phi/(n-1)}F^2]
\nonumber
\eea
where R and $\Phi$ are the usual Ricci scalar and dilaton field respectively, the latter has self-interaction $V(\Phi)$ and non-minimally couples to the electromagnetic field of kinetic energies $F^2$. The quantity $\alpha$ measures the strength of this coupling. Equation of motions following from this action have the form
\bea{feq}
\nonumber
&&\hspace{-5mm}R_{\mu \nu} = \frac{1}{n-1} [4\partial _\mu \Phi \partial _ \nu \Phi +g_{\mu \nu} V(\Phi)]+2e^{-\frac{4\alpha \Phi}{n-1}} [F_{\gamma \mu} F_{\beta \nu} g^{\gamma \beta}                                   \\
&&\hspace{-5mm} \quad \quad-\frac{1}{2(n-1)} g_{\mu \nu}F^2 ] \\
&&\hspace{-5mm} \nabla ^2 \Phi =\frac{\partial _\mu (\sqrt{-g} g^{\mu \nu} \partial _\nu \Phi)}{\sqrt{-g}}=\frac{n-1}{8} \frac{\partial V}{\partial \Phi}-\frac{\alpha}{2} e^{-\frac{4\alpha \Phi}{n-1}} F^2  
\eea
\begin{align}
\label{Maxwell}
&\hspace{-5mm} \nabla _\mu (e^{-\frac{4\alpha \Phi}{n-1}} F^{\mu \nu})=\partial _\mu (\sqrt{-g} e^{-\frac{4\alpha \Phi}{n-1}} F^{\mu \nu} )=0
\end{align}
By adjusting the form of $V(\Phi)$ appropriately, reference \cite{Gao:2004tv, Gao:2005xv} obtains asymptotically AdS black hole solutions of the system analytically,
\bea{LP} 
&&\hspace{-5mm}V=\frac{2\Lambda}{n(n-2+\alpha  ^2)^2} \{-\alpha ^2[(n+1)^2-(n+1)\alpha ^2
\\
&&\quad -6(n+1)+\alpha ^2 +9] \cdot e^{-\frac{4(n-2)\Phi}{(n-1)\alpha}}
\nonumber\\
&&\quad +(n-2)^2 (n-\alpha ^2) \cdot e^{\frac{4\alpha \Phi}{n-1}}
\nonumber\\
&&\quad+4\alpha ^2 (n-1) (n-2) \cdot e^{-\frac{2\Phi (n-2-\alpha ^2)}{(n-1)\alpha}} \}
\nonumber
\eea
\bea{Metric}
ds^2 =-N^2(\rho)f^2(\rho)dt^2+\frac{d\rho ^2}{f^2(\rho)}+\rho ^2R^2(\rho)d\Omega ^2 _{k,n-1}
\eea
where $d\Omega ^2 _{k,n-1}$ is the line element of (n-1)-dimensional hyper surface of constant curvature (n-1)(n-2)k with $k=\pm1, 0$ corresponding to spheric, hyperbolic and plane topology respectively. In the case of only static electric fields occcur in the system, the only nonzero components of $F_{\mu \nu}$ could be obtained from the maxwell equation \eqref{Maxwell}
\bea{solEM}
F_{t\rho} =N(\rho) \frac{q e^{\frac{4\alpha \Phi}{n-1}}}{[\rho R(\rho)]^{n-1}}
\eea
according to the Gauss theorem, the total electric charge of the black hole reads
\bea{BHelec}
Q=\frac{1}{4\pi} \int _{\rho \to \infty}  F_{t\rho}\sqrt{-g} d^{n-1}x=\frac{\Omega _{n-1}}{4\pi}q
\eea
where $\Omega _{n-1}$ is the volume of (n-1)-dimensional unit sphere. With the electrostatic field \eqref{solEM} and metric ansatz \eqref{Metric} substituted into equation \eqref{feq}, reference \cite{Gao:2004tv, Gao:2005xv} obtains
\bea{}
&&\hspace{-6mm}N^2(\rho)=[1-(\frac{b}{\rho})^{n-2}]^{-\gamma (n-3)} 
\label{ttcomponent}\\
&&\hspace{-6mm}f^2(\rho)=  \{[k-(\frac{c}{\rho})^{n-2}][1-(\frac{b}{\rho})^{n-2}]^{1-\gamma (n-2)}
\nonumber\\
&&\hspace{-6mm}\quad -\frac{2\Lambda}{n(n-1)}\rho ^2[1-(\frac{b}{\rho})^{n-2}]^\gamma  \} \cdot [1-(\frac{b}{\rho})^{n-2}]^{\gamma (n-3)}  
\label{rrcomponent}\\
&&\hspace{-6mm}R^2(\rho)=[1-(\frac{b}{\rho})^{n-2}]^\gamma 
\label{anglecomponent}\\
&&\hspace{-6mm}\gamma =\frac{2\alpha ^2}{(n-2)(n-2+\alpha ^2)} 
\\
&&\hspace{-6mm}\Phi(\rho)=\frac{(n-1) \alpha}{2(n-2+\alpha^2)} \ln [1-(\frac{b}{\rho})^{n-2}] 
\label{phifield}
\eea
Here $b$ and $c$ are integration constants with dimension length, they are related with the charge parameter $q$ through
\bea{Qbc}
q^2 =\frac{(n-1)(n-2)^2}{2(n-2+\alpha ^2)} b^{n-2} c^{n-2}
\eea
Reference \cite{Sheykhi:2009pf} shows that for this black hole solution, the Kretschmann scalar $R^{\mu \nu \alpha \beta} R_{\mu \nu \alpha \beta}$ and the Ricci scalar R both diverge at $\rho =b$, thus $\rho =b$ is the location of curvature singularity. On the other hand, it can be easily verified that as $\rho \to \infty$, $N^2(\rho) \to 1, R^2(\rho) \to 1$ and
\begin{align}
f^2(\rho)=\left\{
\begin{array}{l}
1+\frac{\rho ^2}{L^2} -\frac{2\alpha ^2}{1+\alpha ^2} \frac{\rho b}{L^2}+\frac{\alpha ^2(\alpha ^2-1)}{(1+\alpha^2)^2} \frac{b^2}{L^2},n=3
\\
1+\frac{\rho ^2}{L^2}-\frac{\alpha ^2}{2+\alpha^2} \frac{b^2}{L^2},\quad \quad n=4
\\
1+\frac{\rho ^2}{L^2},\quad \quad \quad n\ge 5
\end{array}
\right.
\end{align}
As results, the geometry is asymptotically approximate-AdS in dimensions 3+1 and 4+1, while asymptotically exact AdS in dimenstions $n+1\geqslant 6$. Finally we note that when $\alpha =\gamma =0$ both the dilaton field and it's potential will vanish identically and action $\eqref{EMD}$ will reduce to the usual Einstein-Maxwell ones, the corresponding dilaton black holes also naturally simplifies to the usual Riessner-Nordstr\"om-AdS black holes.

Just as will be shown in the following, radial positions of the black hole horizon defined by $f(\rho _+) =0$ is important for our discussion. However, due to the complicated form of $f(\rho)$, we cannot find explicit expressions for $\rho _+$ through this definition. Alternatively, we choose to use this definition to express parameters $c$ in it in terms of $\rho _+$. 
\bea{repc}
\nonumber
&&c=\rho _+ \cdot\big[1-(\frac{b}{\rho _+})^{n-2}\big]^{\gamma -\frac{1}{n-2}}\cdot\big\{\big[1-(\frac{b}{\rho _+})^{n-2}\big]^{1-(n-2)\gamma} \\
&&\quad \quad -\frac{2\rho ^2 _+[1-(\frac{b}{\rho _+}]^{n-2})^\gamma \Lambda }{n(n-1)}\big\}^{\frac{1}{n-2}}
\eea
 So, in our discussions, the free parameters are chosen as $b$, $\Lambda$, $\rho_+$, $\gamma$ and $n$.

\section{Thermodynamic definitions in the extended phase space
\label{calThermodynamic}}

The black hole mass $M$ is conserver charges associated with time translation symmetries, it can be obtained by the ADM decomposition of metrics on the spacetime boundary. For dilaton black holes of \eqref{Metric}-\eqref{phifield}, ref. \cite{Sheykhi:2009pf, Dayyani:2016gaa, Sheykhi:2008rk} tells us,
\bea{BHmass}
\nonumber
&&M=\frac{(n-1)\Omega _{n-1}\rho _+^{n-2}}{16\pi}[1+\frac{(n-2-\alpha ^2)(\frac{b}{\rho_+})^{n-2}}{n-2+\alpha ^2} \\
&&\quad\quad\quad-\frac{2\rho ^2_+ \Lambda(1-(\frac{b}{\rho_+})^{n-2})^{\frac{n}{n-2}-\frac{2(n-1)}{n-2+\alpha^2}}}{n(n-1)}]
\eea 
In the extended phase space thermodynamic, $M$ has the physical meaning of chemical enthalpy, which is the total energy  of a system including both internal energies $E$ and the $P$-v term, the latter is used to subtract the infinite term caused by the space-integration of cosmological constant\cite{adsLambdaThermo1, Kubiznak1608}. The Hawking temperature can be calculated as 
\beq{}
T=\frac{\sqrt{[N^2(\rho _+) f^2(\rho _+)]'\cdot [f^2(\rho _+)]'}}{4\pi} 
\eeq 
with the definition that $f(\rho _+)=0 $, the above formula can be further translated into
\bea{Temper}
\nonumber
&&\hspace{-10mm}T=\frac{(n-2)\big[1-(\frac{b}{\rho _+})^{n-2}\big]^{\frac{1}{2-n}+\frac{n-1}{n-2+\alpha ^2}}}{4\pi \rho _+} \\
\nonumber
&&\hspace{-10mm}\quad-\frac{\big[1-(\frac{b}{\rho _+})^{n-2}\big]^{\frac{1}{n-2}+\frac{1-n}{n-2+\alpha ^2}}\rho _+ \Lambda}{2n\pi (n-1)(n-2+\alpha ^2)} \times\\
&& \hspace{-10mm}\quad [(n(n-2+\alpha ^2))-2(n-2)(n-1)(\frac{b}{\rho _+})^{n-2}]
\eea
The entropy of the system, according area laws, is just one quarter of the event horizon area \cite{Sheykhi:2009pf}
\bea{entropy}
S=\frac{\Omega _{n-1} b^{n-1} (1-(\frac{b}{\rho _+})^{n-2})^{\frac{(n-1)\gamma}{2}}}{4(\frac{b}{\rho _+})^{n-1}}
\eea
While the electric charge and chemical potentials can be written respectively as
\bea{Qnoc}
\nonumber
&&Q=\frac{(b\rho _+)^{\frac{n-2}{2}}\Omega _{n-1}}{4\sqrt{2} \pi}  \sqrt{\frac{(n-2)^2 (n-1)}{n-2+\alpha ^2}}  
\cdot \Big[1-
\\
&&\quad \quad\frac{2\rho ^2 _+ \Lambda (1-(\frac{b}{\rho_+})^{n-2})^{\frac{n}{n-2}-\frac{2(n-1)}{n-2+\alpha ^2}}}{n(n-1)}\Big]^\frac{1}{2}
\eea 
\beq{elecpoten}
U=A_0 \vert _{\rho \to \infty} - A_0 \vert _{\rho \to \rho _+}
= -\frac{q}{(n-2)\rho _+ ^{n-2}}
\eeq
The last pair of general coordinates for thermodynamic discussions is the pressure and volumes, \beq{pressure}
P=-\frac{\Lambda}{8\pi},~V=(\frac{\partial M}{\partial P})_{S,Q} 
\eeq 
with eqs. \eqref{BHmass} and \eqref{pressure}, we can express $V$ in terms of $b$, $\rho$, $\alpha$ et al as follows
\bea{therVOL}
&&\hspace{-5mm}V={(n-1)\rho_+^n\big[1-(\frac{b}{\rho _+})^{\frac{2(n-1)\alpha ^2}{(n-2)(n-2+\alpha ^2)}}\big]} \\
&&\hspace{-5mm}\quad \quad \cdot  \frac{(n-2+\alpha ^2)-(n-2)(\frac{b}{\rho _+})^{n-2}}{6(n-2+\alpha ^2)(1-(\frac{b}{\rho _+})^{n-2})}
\eea
In the case of $n=3,\alpha =0$, this reduces to the volume of 4-dimensional RN-AdS black holes naturally.

With thermodynamic quantities presented above, we check that in the  extended phase space method, the first law of black hole thermodynamics is indeed properly holden. In fact, by the link rule of differentiations [Note the extra minus sign on the right hand most expression, which is caused by the implicit function derivative $Q(b,\rho_+)=Q_0$], 
\bea{}
\nonumber
\nonumber
&&\hspace{-5mm}(\frac{\partial M}{\partial S})_{P,Q}= (\frac{\partial M}{\partial \rho _+} \cdot \frac{\partial \rho _+}{\partial b}+\frac{\partial M}{\partial b})_{P,Q}/(\frac{\partial S}{\partial \rho _+} \cdot \frac{\partial \rho _+}{\partial b}+\frac{\partial S}{\partial b})_{P,Q} \\
\nonumber
&&\hspace{-5mm}(\frac{\partial \rho _+}{\partial b})_{P,Q} =-[(\frac{\partial Q}{\partial b})/(\frac{\partial Q}{\partial \rho _+})]_P \\
\nonumber
&&\hspace{-5mm}(\frac{\partial M}{\partial Q})_{P,S}= (\frac{\partial M}{\partial \rho _+} \cdot \frac{\partial \rho _+}{\partial b}+\frac{\partial M}{\partial b})_{P,S}/(\frac{\partial Q}{\partial \rho _+} \cdot \frac{\partial \rho _+}{\partial b}+\frac{\partial Q}{\partial b})_{P,S} \\
\nonumber
&&\hspace{-5mm}(\frac{\partial \rho _+}{\partial b})_{P,S} =-[(\frac{\partial S}{\partial b})/(\frac{\partial S}{\partial \rho _+})]_P
\eea
and some tedious but straightforward calculation with $\eqref{BHmass},\eqref{Temper},\eqref{entropy}$, we can easily prove that                                            
\beq{dMds}
(\frac{\partial M}{\partial S})_{P,Q}=T,~(\frac{\partial M}{\partial Q})_{P,S}=U  
\eeq
Thus, these thermodynamics quantities indeed satisfy the first law of black hole thermodynamics in the extended phase space method
\bea{TherFlaw}
dM=TdS+UdQ+VdP 
\eea
According to eqs. \eqref{Temper} and \eqref{pressure}, the system's equation of state can be written as
\bea{stateEQ}
\nonumber
&&\hspace{-5mm}P=\frac{n(n-1)(n-2+\alpha ^2)(1-(\frac{b}{\rho _+})^{n-2})^{\frac{1}{2-n}+\frac{n-1}{n-2+\alpha ^2}}}{4\rho _+ (n(n-2+\alpha ^2)-2(n-2)(n-1)(\frac{b}{\rho _+})^{n-2})} \times \\
&&\hspace{-5mm}\quad \quad [t-\frac{(n-2)(1-(\frac{b}{\rho_+})^{n-2})^{\frac{1}{2-n}+\frac{n-1}{n-2+\alpha ^2}}}{4\pi \rho _+}]
\eea
with parameters $b$, $\rho_+$ and $\alpha$ et al related with the specific volume of the system as follows
\bea{Svol}
&&\hspace{-10mm}v=\frac{4\rho _+ (n(n-2+\alpha ^2)-2(n-2)(n-1)(\frac{b}{\rho _+})^{n-2})}{n(n-1)(n-2+\alpha ^2)(1-(\frac{b}{\rho _+})^{n-2})^{\frac{1}{2-n}+\frac{n-1}{n-2+\alpha ^2}}}
\eea

As in the usual thermodynamic theories, for the charged dilaton black holes, the Gibbs free energy and specific heat of the system are defined as follows
\bea{}
\label{FreeEnergy}
&&G=M-TS\\
\label{SpeHeat}
&&C_p=-T\frac{\partial ^2 M}{\partial T ^2}|_{p~\mathrm{fixed}}
\eea   
In canonical ensembles, we will fix the $T$, $Q$ parameter when exploring physics in the  $P$-v and $G-P$ plane; we will fix the $P$, $Q$ parameter when exploring the $C_p$-$T$ physics. The $P$-v relation in canonical ensembles is derived from eqs.\eqref{Temper}, \eqref{Qnoc}, \eqref{pressure} and \eqref{Svol}, which will be denoted as 
\bea{}
\label{TabrhP}
T=T(b,\rho _+,P,\alpha,n),
\\
\label{QabrhP}
Q=Q(b,\rho _+,P,\alpha,n),
\\
\label{vabrhp}
\mathrm{v}=\mathrm{v}(b,\rho _+,P,\alpha,n).
\eea   
Fixing the value of $Q=Q_0$ allows us to express $P$ as follows
\bea{PfromQ}
P=P(b,\rho _+,Q_0,\alpha,n).
\eea   
Substituting this expression into \eqref{TabrhP} with fixed $T=T_0$ will yield us 
\bea{rhabTQ}
\rho _+=\rho _+(b,T_0,Q_0,\alpha,n)
\eea   
Finally, substituting eq.\eqref{rhabTQ} back into \eqref{PfromQ} and into $\eqref{vabrhp}$, we will get equation of states with fixed temperature and electric charges 
\bea{}
\label{PabTQ}
P=P(b,T_0,Q_0,\alpha,n) \\
\label{vabTQ}
\mathrm{v}=\mathrm{v}(n,\alpha,b,T_0,Q_0)
\eea

Similiarly, the $C_p$-$T$ relation in canonical ensembles is also obtained from eqs.\eqref{FreeEnergy}, \eqref{Qnoc}, \eqref{pressure} and \eqref{Temper} with fixed pressures $P=P_0$
\bea{}
\label{reQabrhP}
Q=Q(b,\rho _+,P_0,\alpha,n)\\
\label{CpabrhP}
C_p=C_p(b,\rho _+,P_0,\alpha,n) \\
\label{TpabrhP}
T=T(b,\rho _+,P_0,\alpha,n)
\eea
Fixing electric charges $Q=Q_0$ in \eqref{reQabrhP} and solve out from it $\rho _+$, we will have
\bea{rhabPQ}
\rho _+=\rho _+(b,P_0,Q_0,\alpha,n).
\eea
Substituting this result back into eqs\eqref{CpabrhP} and \eqref{TpabrhP}, we will get the equation of state with fixed pressure and electric charges as follows
\bea{}
\label{CpabPQ} 
C_p=C_p(b,P_0,Q_0,\alpha,n)\\
\label{TabPQ}
T=T(b,P_0,Q_0,\alpha,n)
\eea
Due to the complicated form of formulas involved in these thermodynamical quantities, we will not (sometimes cannot) write out the concrete form of \eqref{PabTQ} and  \eqref{CpabPQ} here. However, they indeed play key roles in our numeric explorations in the following section.  

\section{Phase structure and thermodynamical analysis \label{BHThermodynamic}}

From actions \eqref{EMD}, we easily see that different $\alpha$ corresponds to different gravitation theories. For example, $\alpha =0$ corresponds to the usual Einstein-Maxwell theory; $\alpha=1$ corresponds to the Einstein-Maxwell-Dilaton theory that follow from low energy limit string theories in the Einstein frame. From the viewpoint of AdS/CFT correspondence, different $\alpha$ will correspond to different thermodynamic systems. For this reason, we naturally expect that different phase structure or transition patterns for different $\alpha$'s will occur in the extended phase space thermodynamics. From another point of view, the value of $\alpha$ characterizes the strength of string corrections in the full theory. By comparing phenomenas in the nonzero-$\alpha$ theories with those in the $\alpha=0$ one, we have chances to see string effects of the theory relative the simple Einstein-Maxwell gravitations intuitively.        

\begin{figure}[ht]
\begin{center}
\includegraphics[scale=0.38]{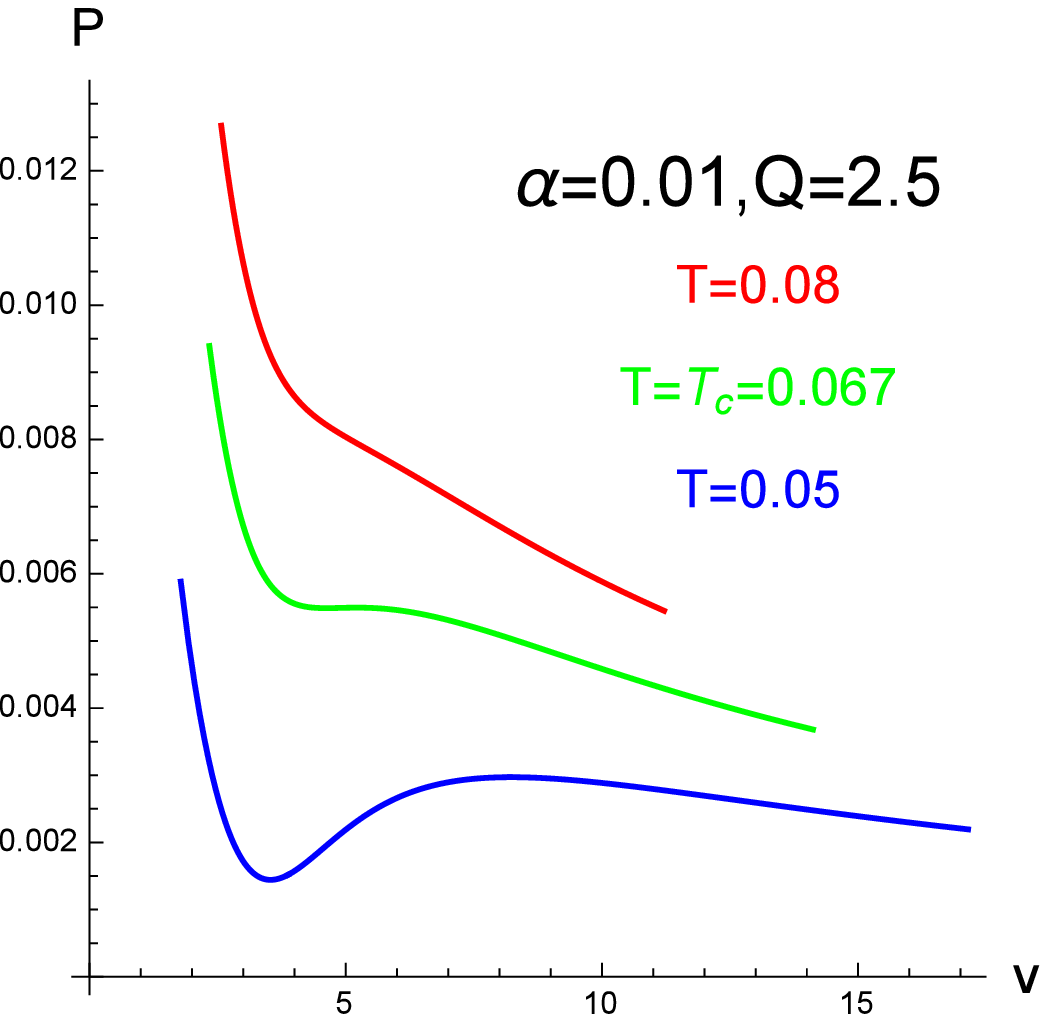}
\includegraphics[scale=0.38]{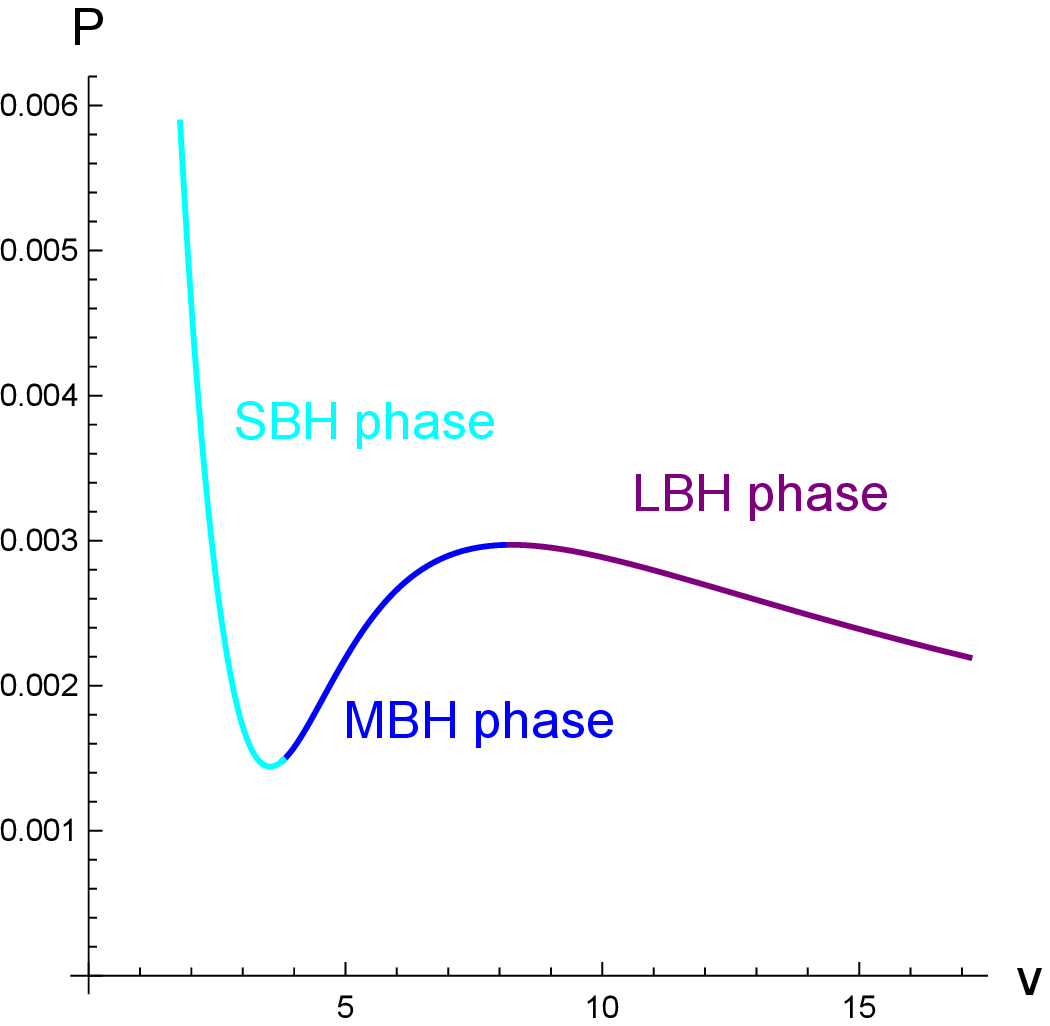}
\caption{(color online). The P-V line of 5-dimensional charged Dilaton AdS black holes with Q=2.5, $\alpha =0.01$. The left panel is the isotherm line for temperatures being higher than(red), equal to(green) and lower than(blue) the critical temperature $T_c$ respectively. The right panel is the magnifying of isotherm lines of $T<T_c$. The three phase, SBH, MBH and LBH phases in this panel are denoted by cyan, blue, and purple lines respectively.}
\label{PVwithalpha0}
\end{center}
\end{figure}

\begin{figure}[ht]
\begin{center}
\includegraphics[scale=0.38]{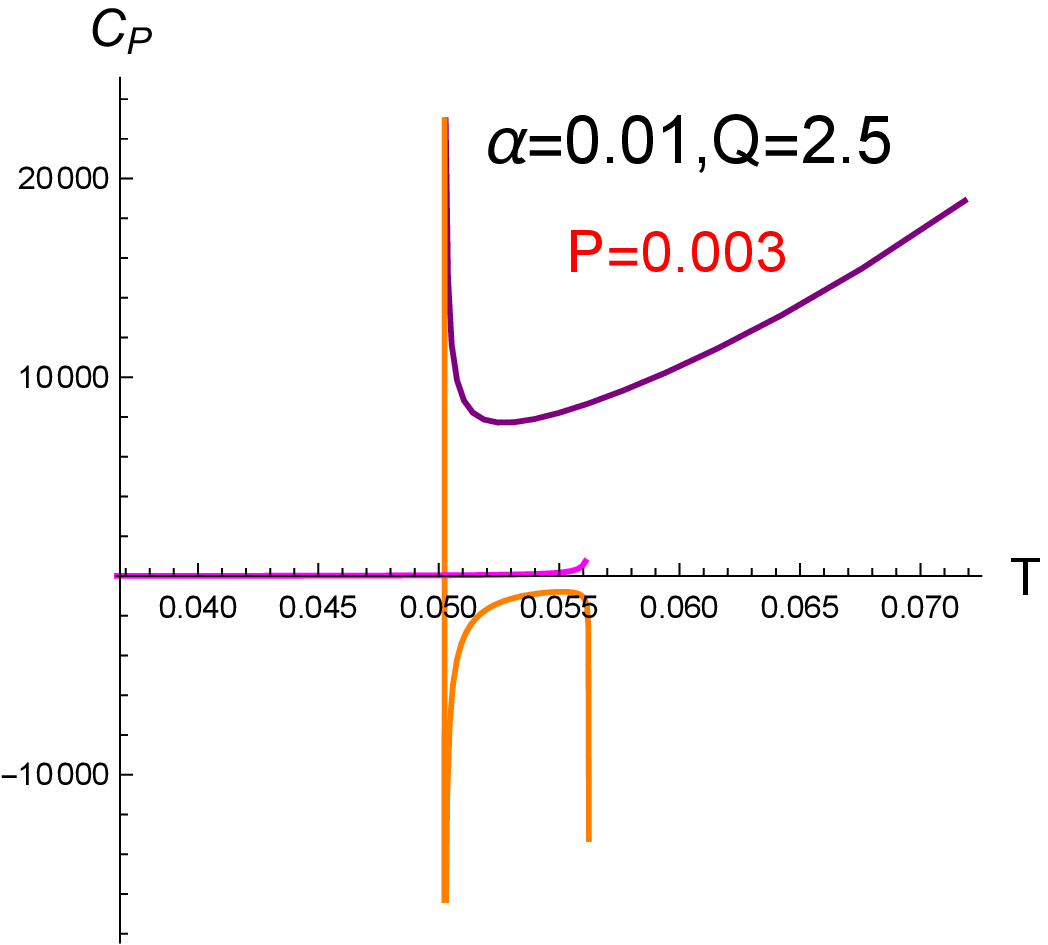}
\includegraphics[scale=0.38]{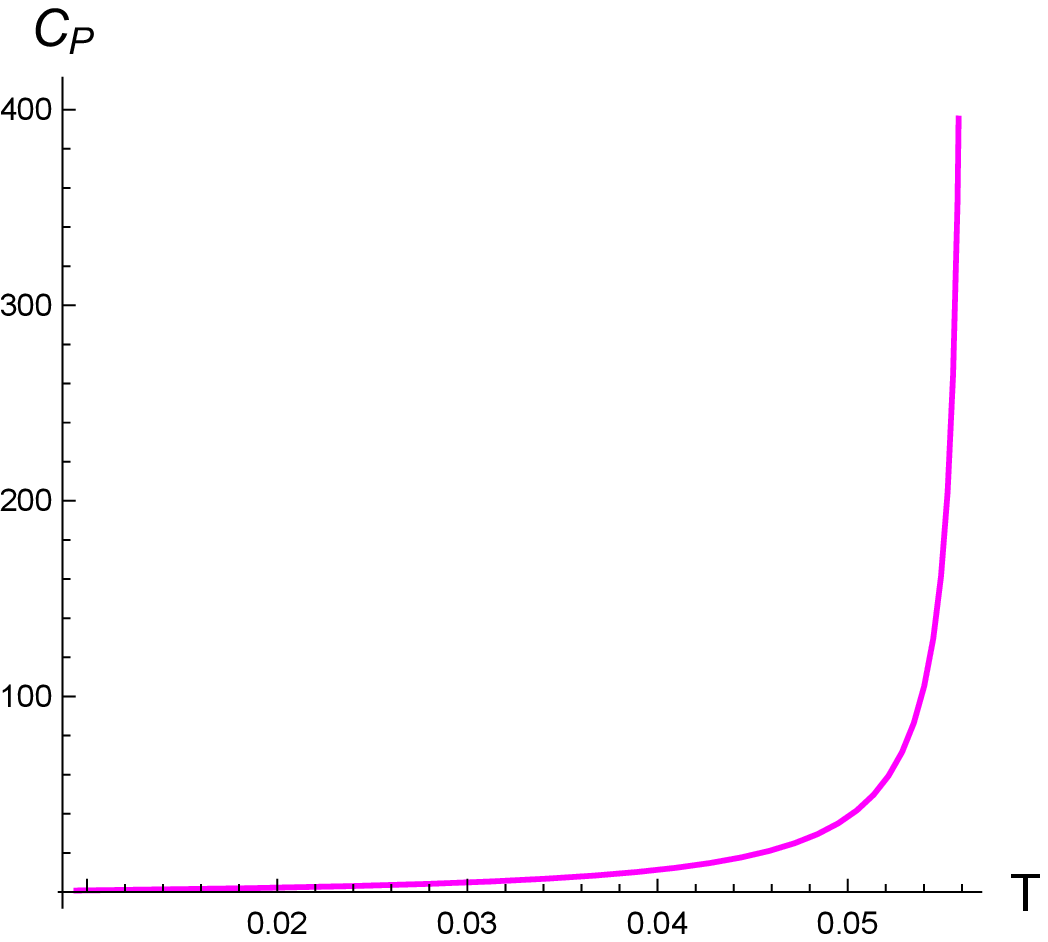}
\\
\includegraphics[scale=0.38]{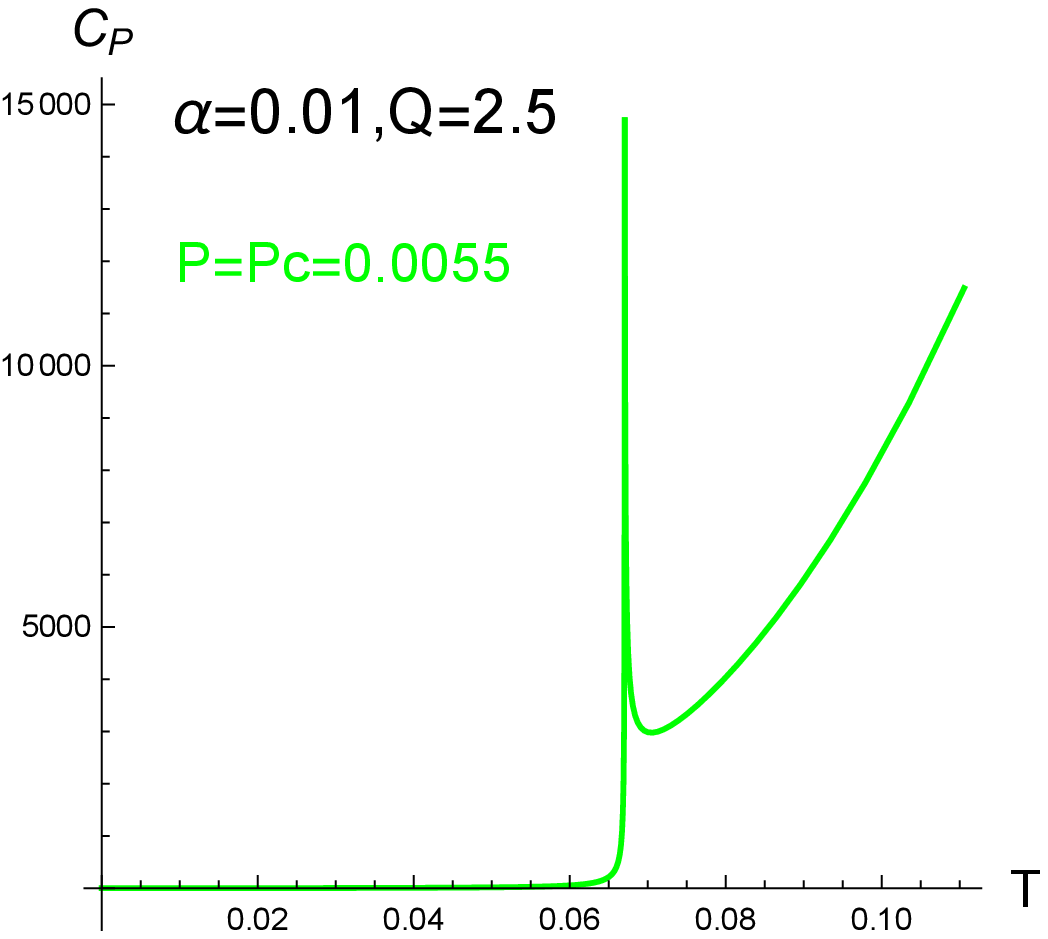}
\includegraphics[scale=0.38]{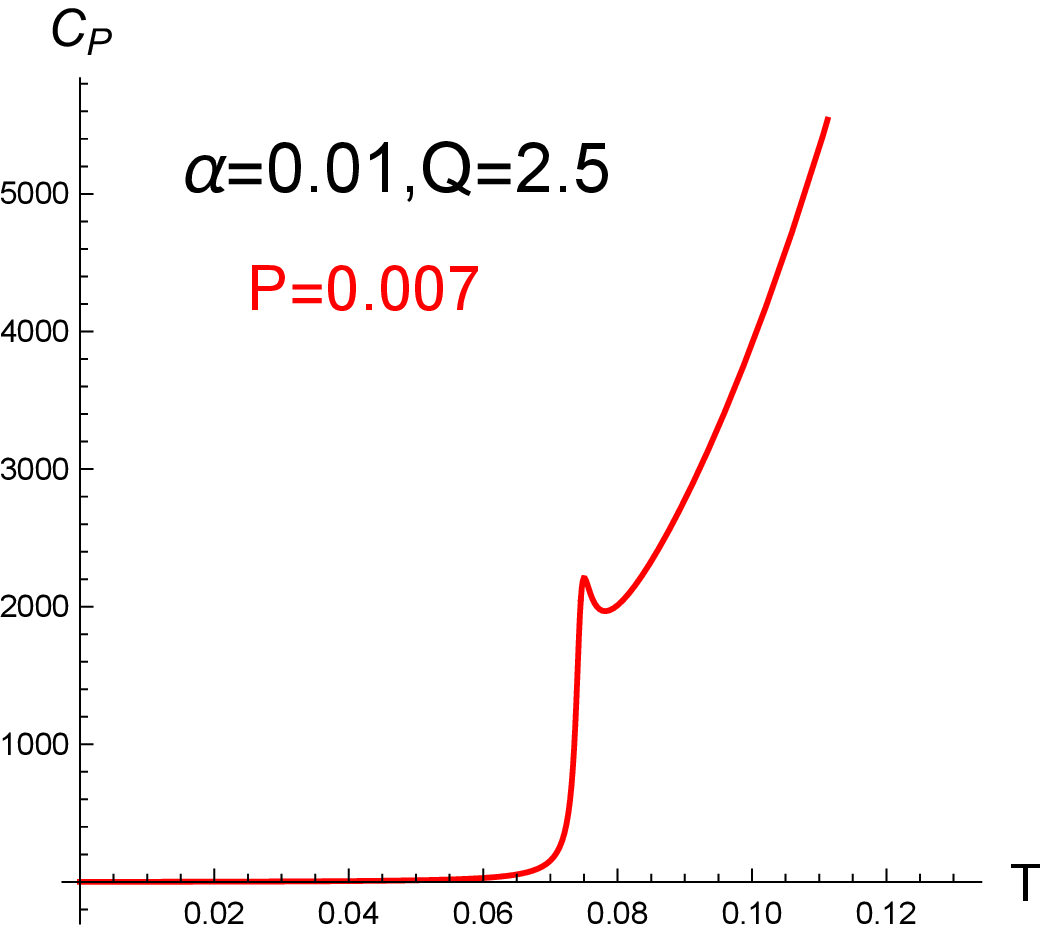}
\caption{(color online). $C_p$-$T$ lines of 5-dimensional charged AdS Dilaton black holes with Q=2.5, $\alpha =0.01$. The top-left is for cases with $P<P_c$, the corresponding $P$-v line is the right panel of FIG.\ref{PVwithalpha0}. The top right is the magnifying of the top-right. The bottom-left and bottom-right are cases with pressures being equal to (green) and less than (blue) the critical value $P_c$ respectively.}
\label{CpTwithalpha0fixP1toP3}
\end{center}
\end{figure}

\begin{figure}[ht]
\begin{center}
\includegraphics[scale=0.38]{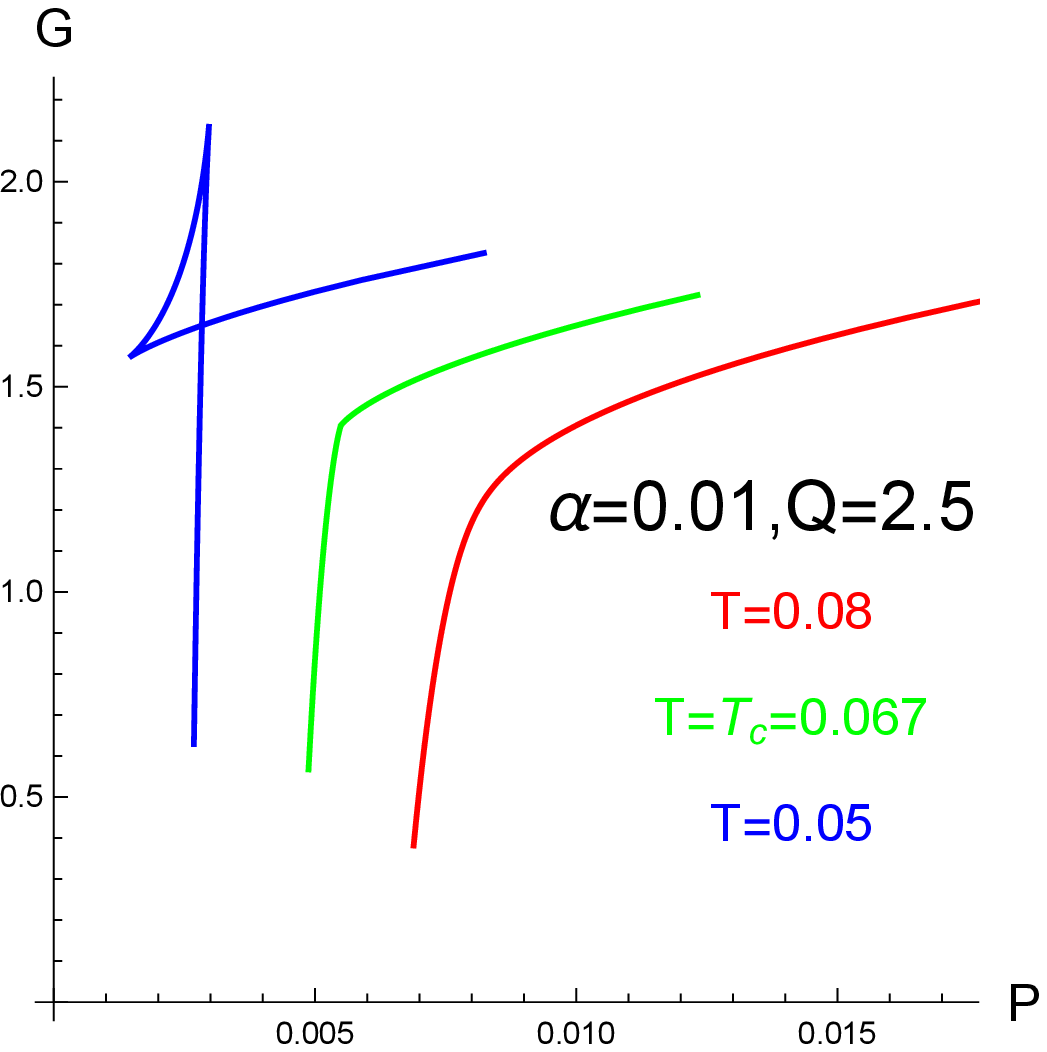}
\includegraphics[scale=0.38]{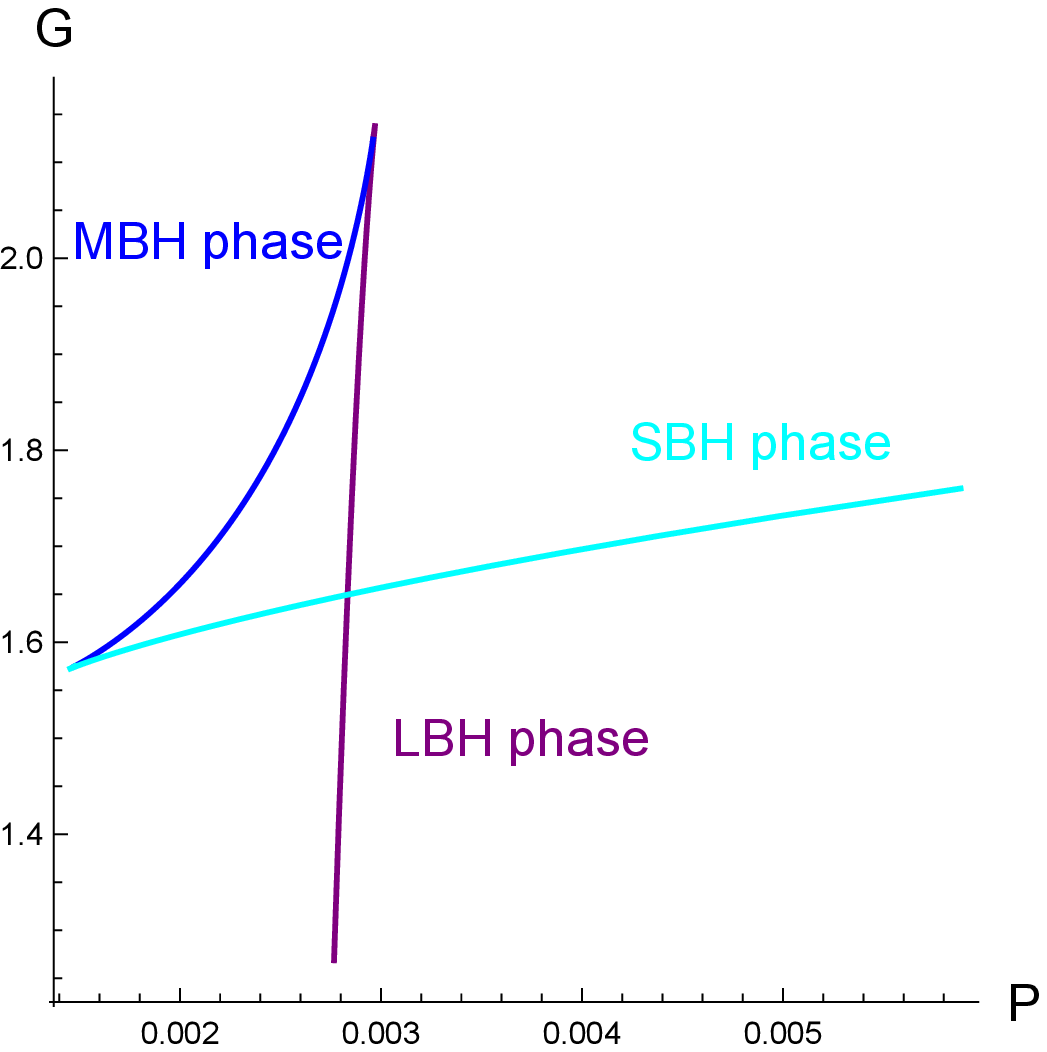}
\caption{(color online). $G$-$P$ lines of 5-dimensional charged dilaton AdS black hole with $Q=2.5$, $\alpha =0.01$, at temperatures being lower than(blue), equal to(green) and higher than(red) the critical temperature $T_c$. The right panle is the magnification of the lowest-temperature line in the left part.}
\label{GPwithalpha0}
\end{center}
\end{figure}

In the small $\alpha$ limit, the action of the system reduces to
\begin{align*}
S \sim \frac{1}{16\pi} \int d^{n+1} x \sqrt{-g} [R-2\Lambda -F^2],
\end{align*}
the charged dilaton black hole will reduce to the AdS-RN one correspondingly. Reference \cite{Kubiznak1205} shows that the van der Waals like critical phenomenon exists in the RN-AdS black holes. So we naturally expect similar phenomena will occur in the charged dilaton black holes, especially for small $\alpha$ cases. To see these phenomena, we plot the $P$-v and $G$-$P$ curves in the canonical ensemble at different temperatures with varying temperature for small $\alpha$ and a fixed $Q$ in FIG.\ref{PVwithalpha0}-\ref{GPwithalpha0} respectively.              

From FIG.\ref{PVwithalpha0}, we easily see that the charged dilaton black hole has unique phase in temperatures above some critical one. While in temperatures lower than that critical value, three phases are possible, that is, large black hole (LBH hereafter), small black hole(SBH), and the middle black hole (MBH) phases. From FIG.\ref{CpTwithalpha0fixP1toP3}, we see that both the LBH and SBH phases have positive specific heat, so both of them are thermodynamically stable. As comparisons, the MBH phase is unstable. From FIG.\ref{GPwithalpha0}, we  see that a first-order phase transition could occur as one varies pressures of the system at temperature lower than the critical value. This is a transition between the SBH and LBH phase and has the same feature as those occur in the Van der Waals liquid-gas system. Although in FIGs.\ref{PVwithalpha0}-\ref{GPwithalpha0}, what we display are only for parameters with $\alpha=0.01$, we note that similar phenomenas always occur as $\alpha<1$. This implies that the stringy or dilaton corrections to the Einstein gravity is not so remarkable when the coupling constant $\alpha$ is less than 1, from thermodynamic viewpoints.   
\begin{figure}[ht]
\begin{center}
\includegraphics[scale=0.23]{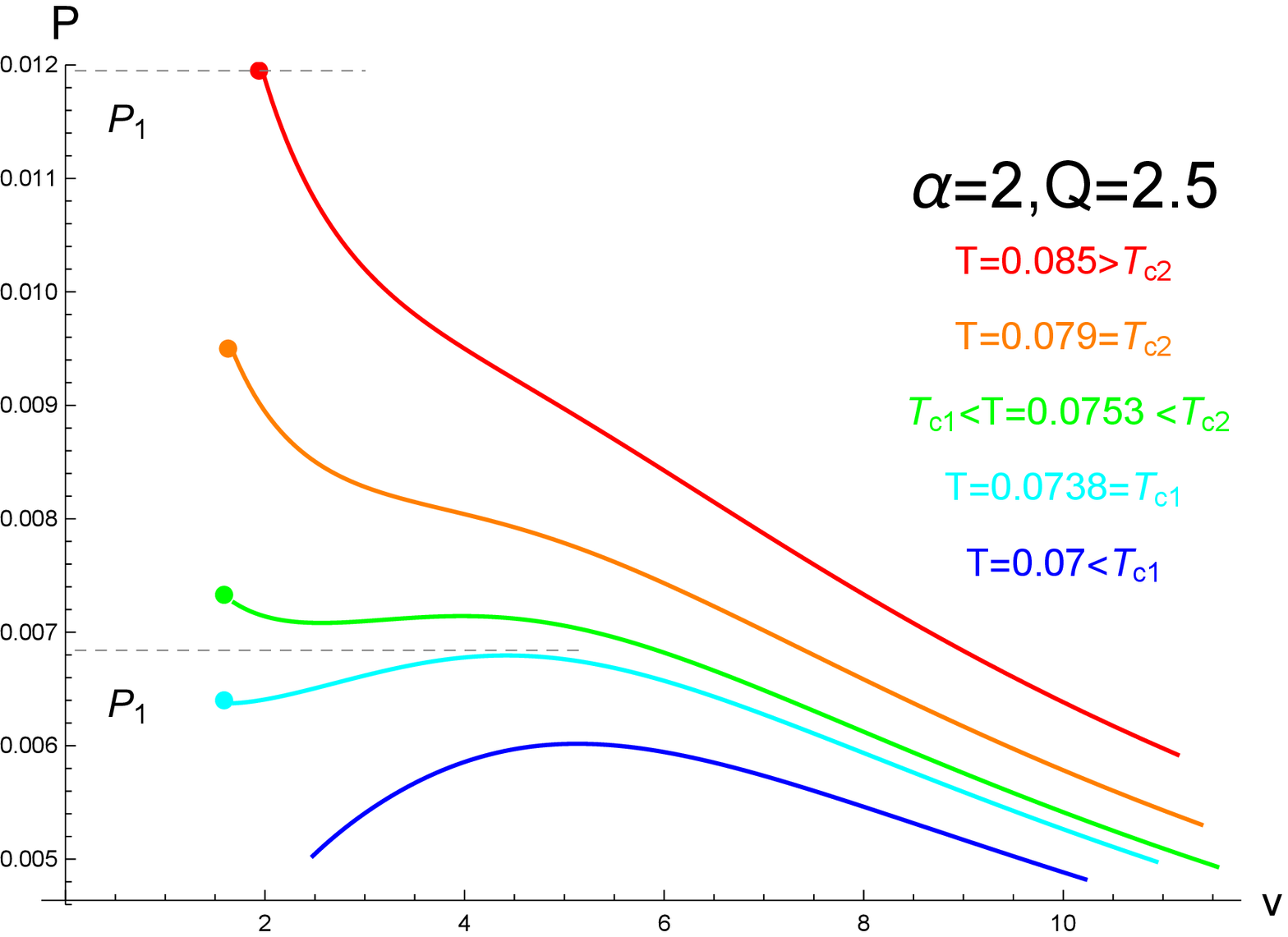}
\includegraphics[scale=0.23]{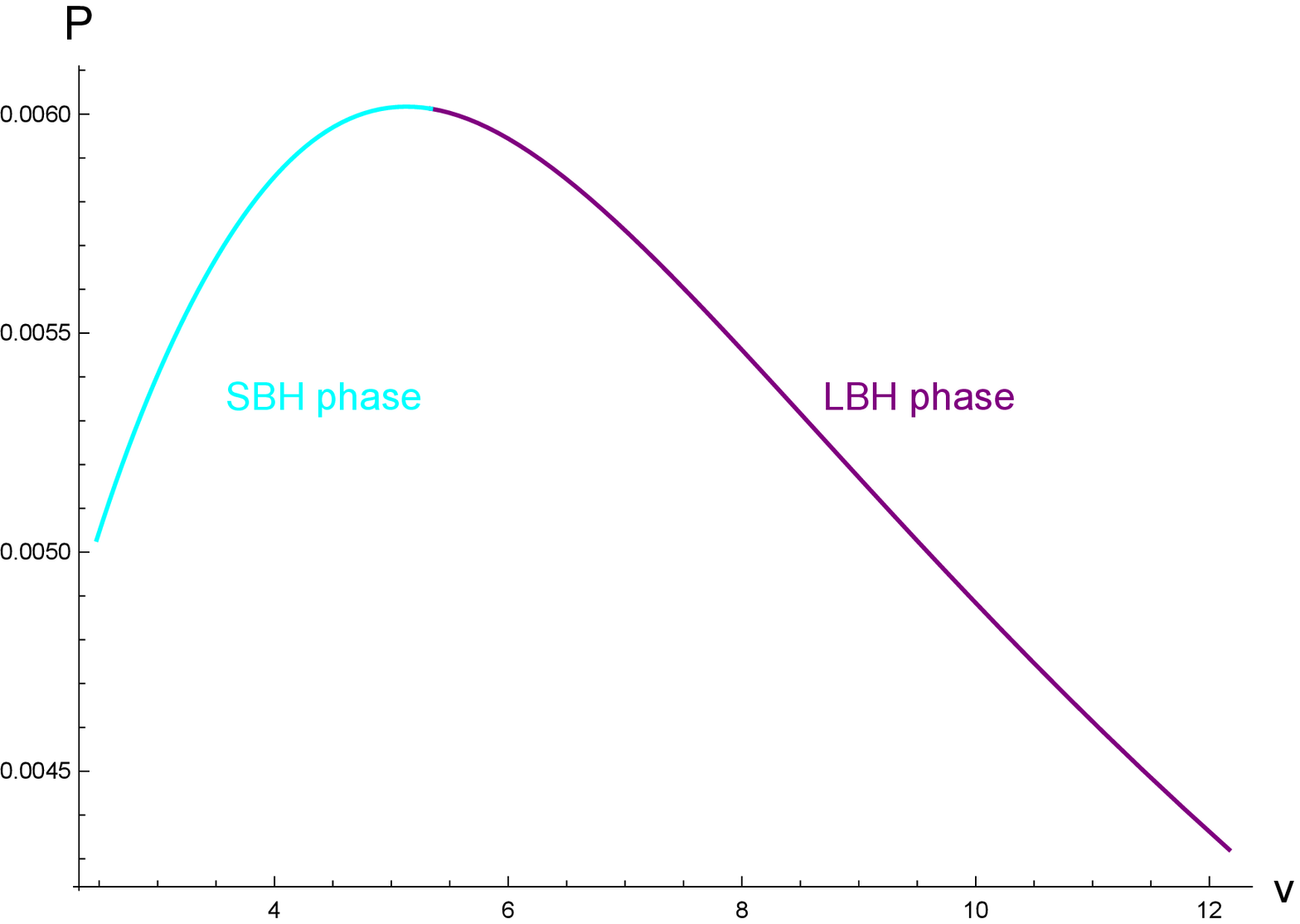}
\caption{(color online) The $P$-v diagram of 5-dimensional charged dilaton AdS black hole with Q=2.5, $\alpha =2$. The left panel is isotherm lines with different temperatures. Two critical temperature case $T_{c1}$ and $T_{c2}$ are displayed in this figure. Acrossing $T_{c1}$, two phase line becomes three phase, while across $T_{c1}$, three phase line becomes one phase. The right panel is the magnification of the isotherm line of $T<T_{c1}$}
\label{PVwithalpha2}
\end{center}
\end{figure}        
\begin{figure}[!ht]
\begin{center}
\includegraphics[scale=0.4]{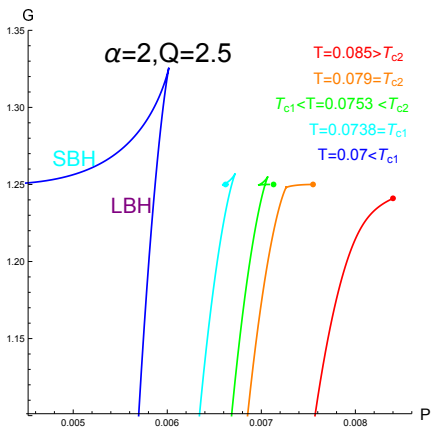}
\caption{(color online) The $G$-$P$ diagram of 5-dimensional charged dilaton AdS black holes whose $P$-v diagram are displayed in FIG.\ref{PVwithalpha2}.}
\label{GPwithalpha2}
\end{center}
\end{figure}         

As the value of $\alpha$ goes larger than $1$, new phenomenas beyond the Van der Waals liquid-gas like transition occur. They are displayed in Fig.\ref{PVwithalpha2}-\ref{GPwithalpha2}, where $\alpha=2$ is chosen as representations. Comparing with those revealed in FIG.\ref{PVwithalpha0}, a significant new point for the $\alpha=2$ case is that there exists a 0-th order phase transition in the $P$-v diagram. As displayed in FIG.\ref{PVwithalpha2}, for temperature $T>T_{c1}$, as the pressure increases and exceeds an upper bound $P_1$, the charged dilaton black hole disappears as an allowed solution to the action system. This is the 0-th order phase transition on pressure $p_1$. While as $T<T_{c1}$, a new type of $P$-v matter state, the blue line in the figure, occurs. This state is featured by the lack of higher-pressure region. In gravitation side, only two kinds of black holes, small (unstable) and large (stable), dual to this matter state exist. Recall that in the Van de Waals like liquid-gas state, three kinds of dual black holes are there. FIG.\ref{GPwithalpha2} displays the Gibbs free energy of these black holes correspondingly. From the figure we easily see that in the new $P$-v state, the LBH phase has more lower free energy is thus thermal-dynamically favored. On the other hand, by comparing thermal quantities in $\alpha=2$ black holes with those in $\alpha\approx0$ ones, we see that it is the string or dilaton correction effects that brings the system this 0-th order phase transition and makes the phase structure of the system richer and complicated.        

\begin{figure}[!ht]
\begin{center}
\includegraphics[scale=0.35]{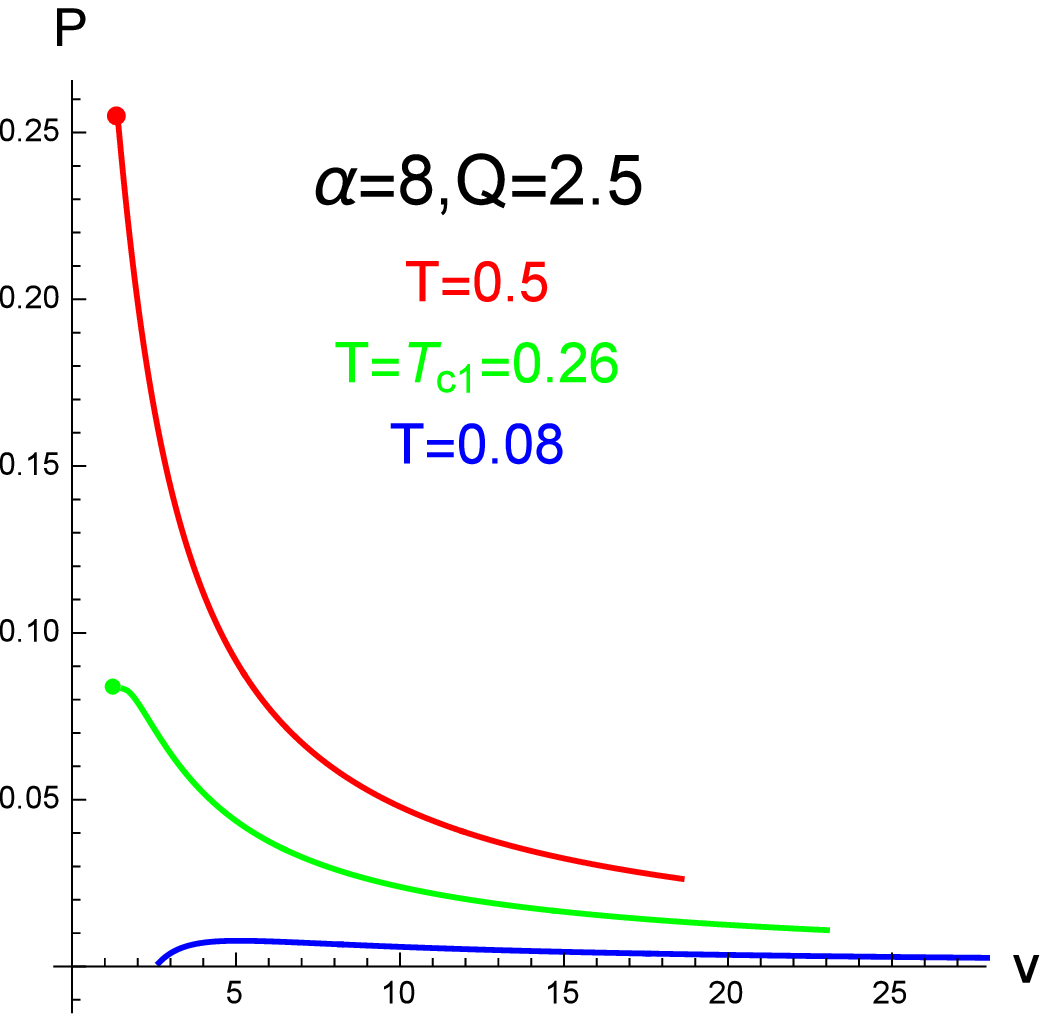}
\includegraphics[scale=0.35]{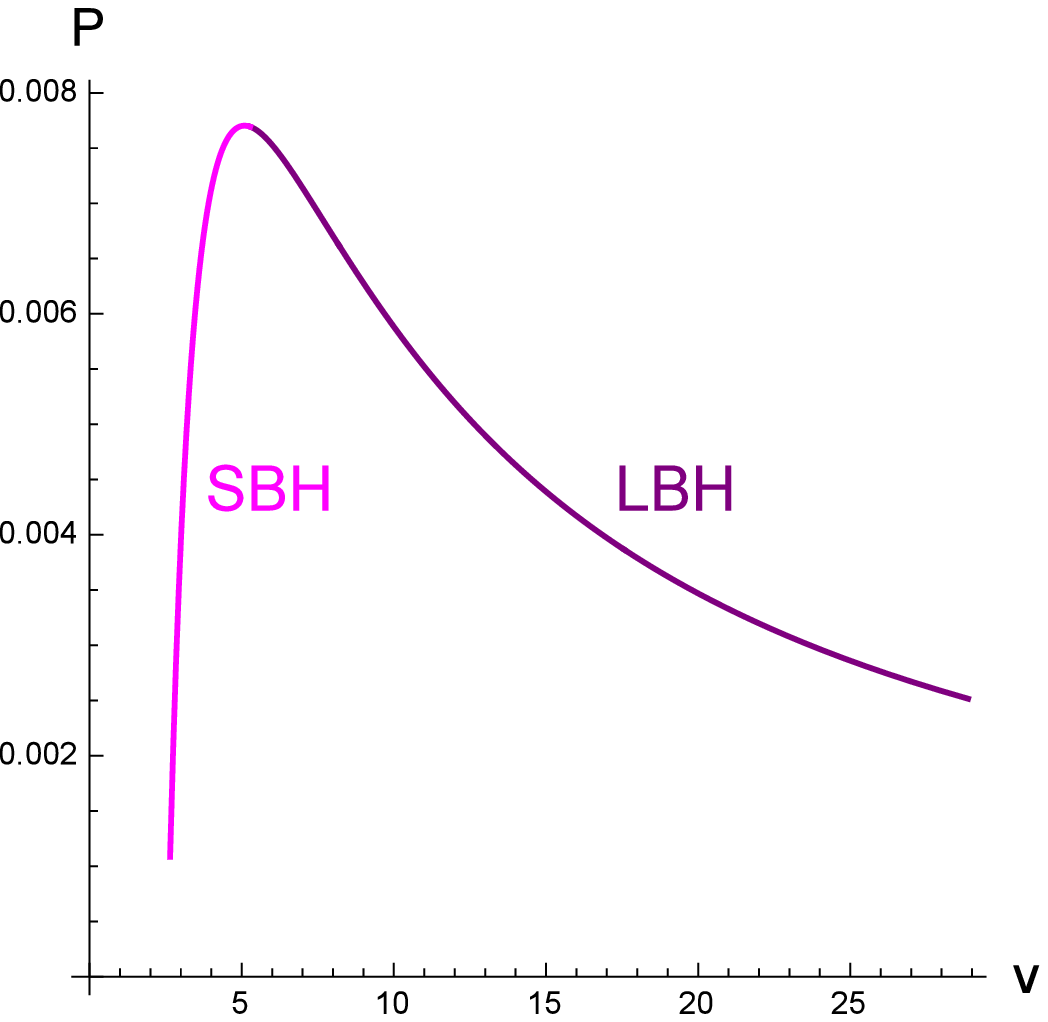}
\caption{(color online) The $P$-v diagram of 5-dimensional charged Dilaton AdS black hole with $Q=2.5$, $\alpha=8$, at different temperatures. The right panel is the magnification for the isotherm line with $T<T_{c1}$ from the left. }
\label{Pvwithalpha8}
\end{center}
\end{figure}
\begin{figure}[!ht]
\begin{center}
\includegraphics[scale=0.35]{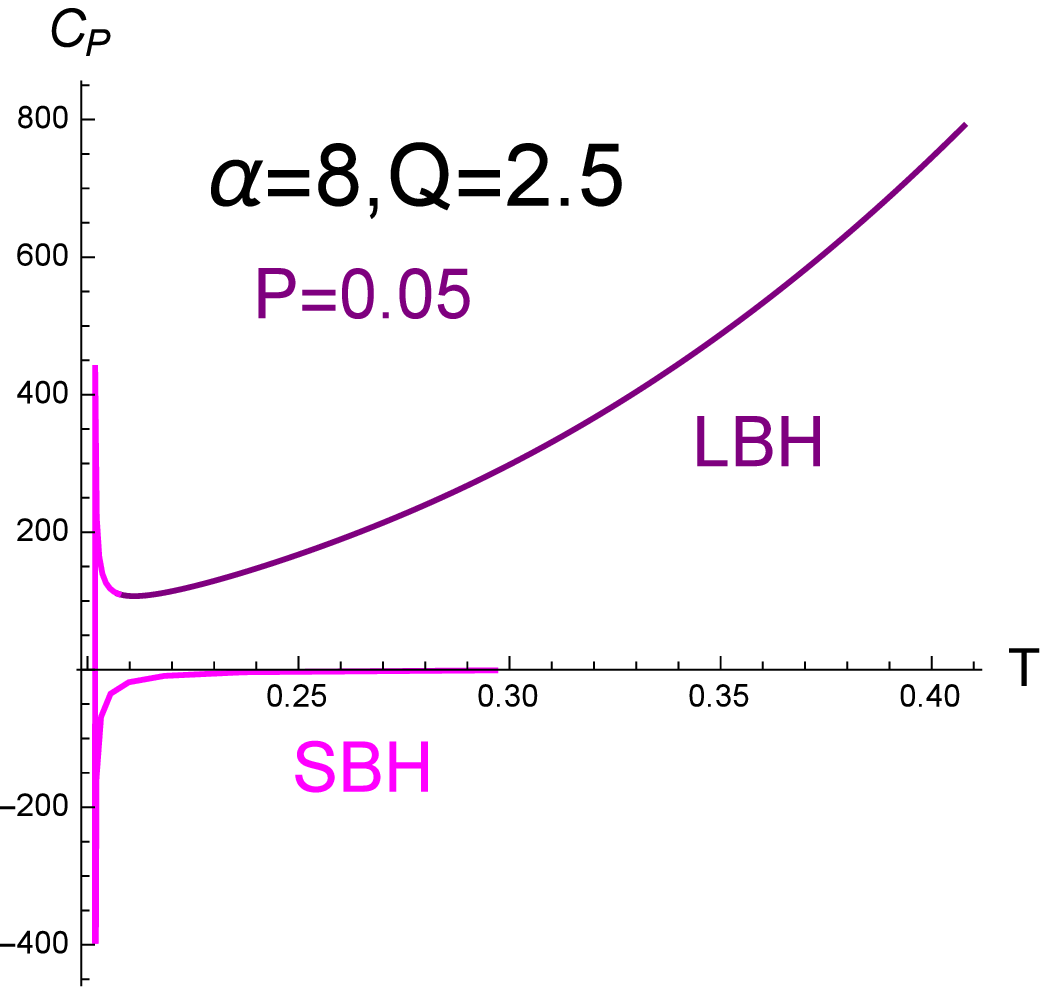}
\includegraphics[scale=0.35]{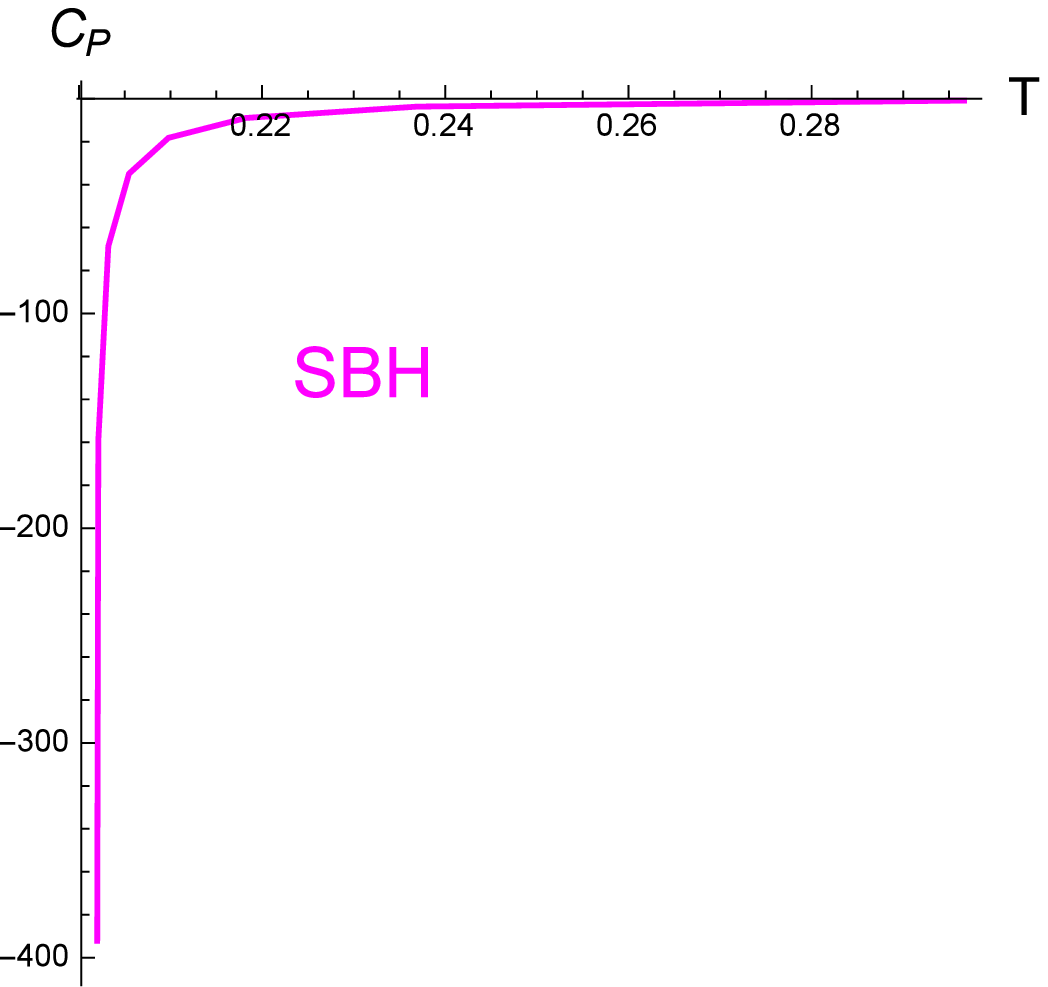}\\
\includegraphics[scale=0.35]{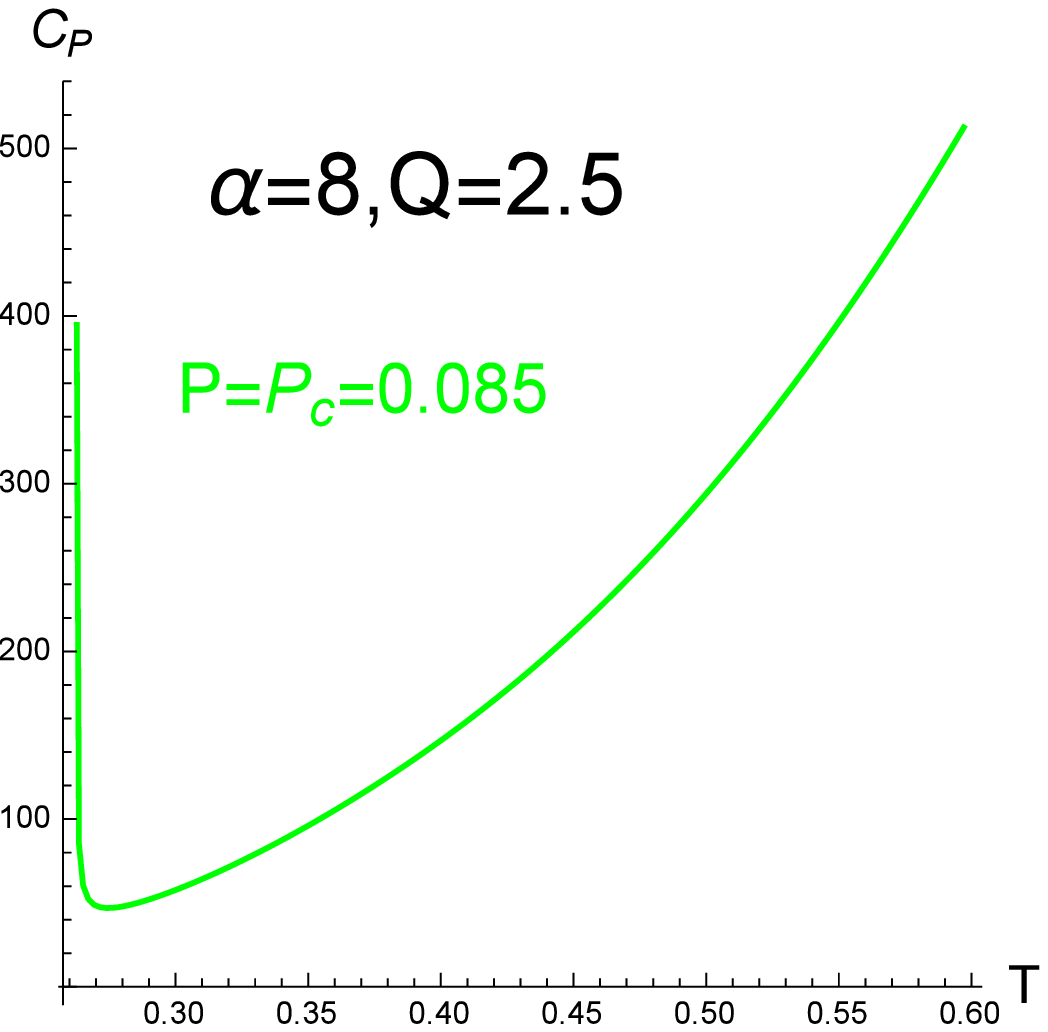}
\includegraphics[scale=0.35]{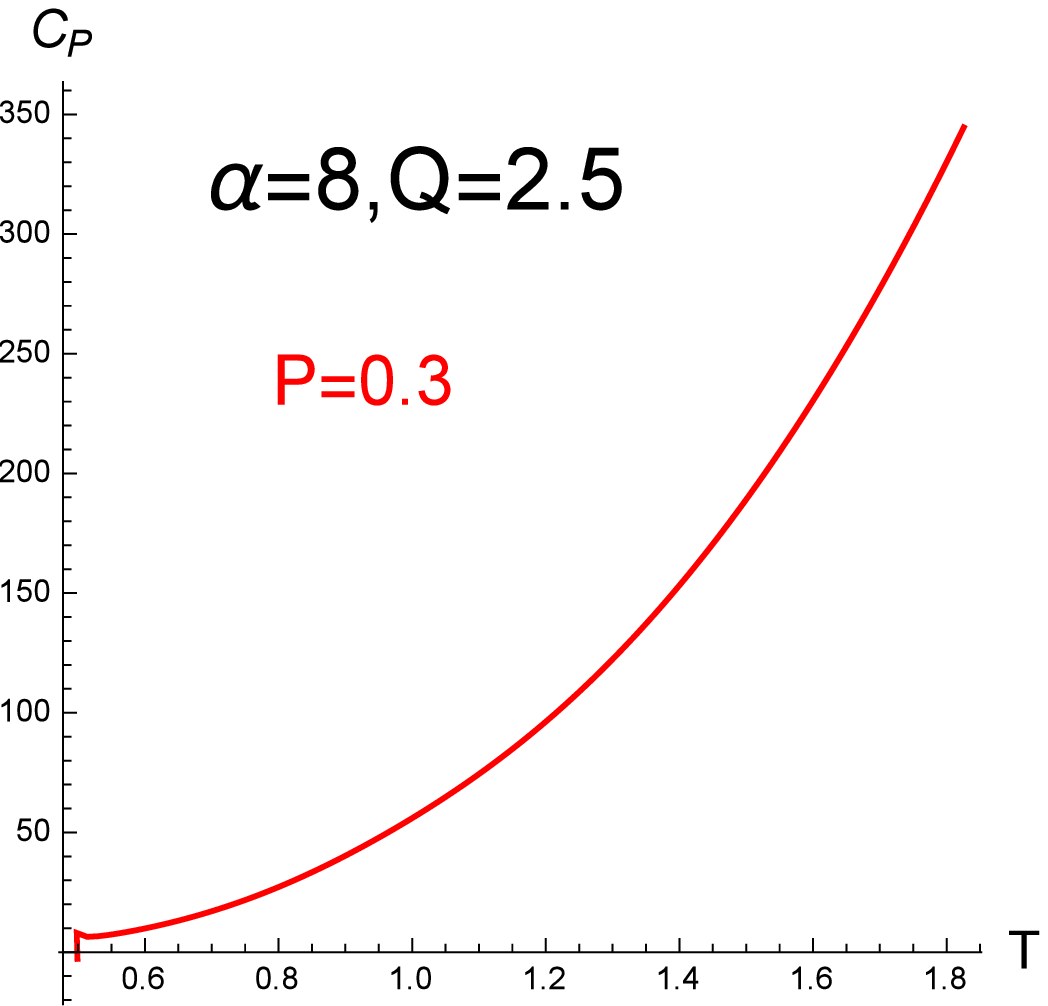}
\caption{(color online) The $C_p$-$T$ diagram of 5-dimensional charged AdS dilaton black holes. Parameters in this figure are corresponding with the left panel of Fig.\ref{Pvwithalpha8}.  The top-right part is magnifications of the SBH phase line in the top right part, which displays the $C_p$-$T$ line when $P<P_c$. The bottom-left and bottom-right part are $C_p$-$T$ lines for pressures equal to(green), higher than(red) the critical pressure $P_c$ respectively.  }
\label{CpTwithalpha8}
\end{center}
\end{figure}
As $\alpha$ goes further large and beyonds some critical value $\alpha_1$, the van der Waals like phase structure begins to shrink from the $P$-v diagram and disappears completely at another critical $\alpha_2$. Between $\alpha_1$ and $\alpha_2$ the $P$-v diagram behaves like that in FIG.\ref{PVwithalpha2}. In the current parameter case $d=5,Q=2.5$, the value of $\alpha_1\approx1.3$, $\alpha_2\approx 2.0$. Their concrete value does not matter in what followings.  FIG.\ref{Pvwithalpha8}-\ref{CpTwithalpha8} displays the $P$-v and $C_p$-$T$ diagram of the system for $\alpha=8$, which is obviously larger than $\alpha_2$. The variation trends of these diagram will not change qualitatively any more as we increase the value of $\alpha$ further. From FIG.\ref{CpTwithalpha8} we easily see that the LBH phase has positive specific heats, while the SBH phase, negative ones. So the former is thermodynamically favored in canonical ensembles. More importantly, such stable phases continue to exist regardless how large $\alpha$ is. So, during the process of Van der waals like liquid-gas phase structures' disappearing from FIG.\ref{PVwithalpha2}, it is the SBH phase that disappears little by little when $\alpha$ goes larger than $\alpha _1$. The LBH phase continues to exist and to be stable in the new P-v state of Fig.\ref{Pvwithalpha8}.

\begin{figure}[!ht]
\begin{center}
\includegraphics[scale=0.41]{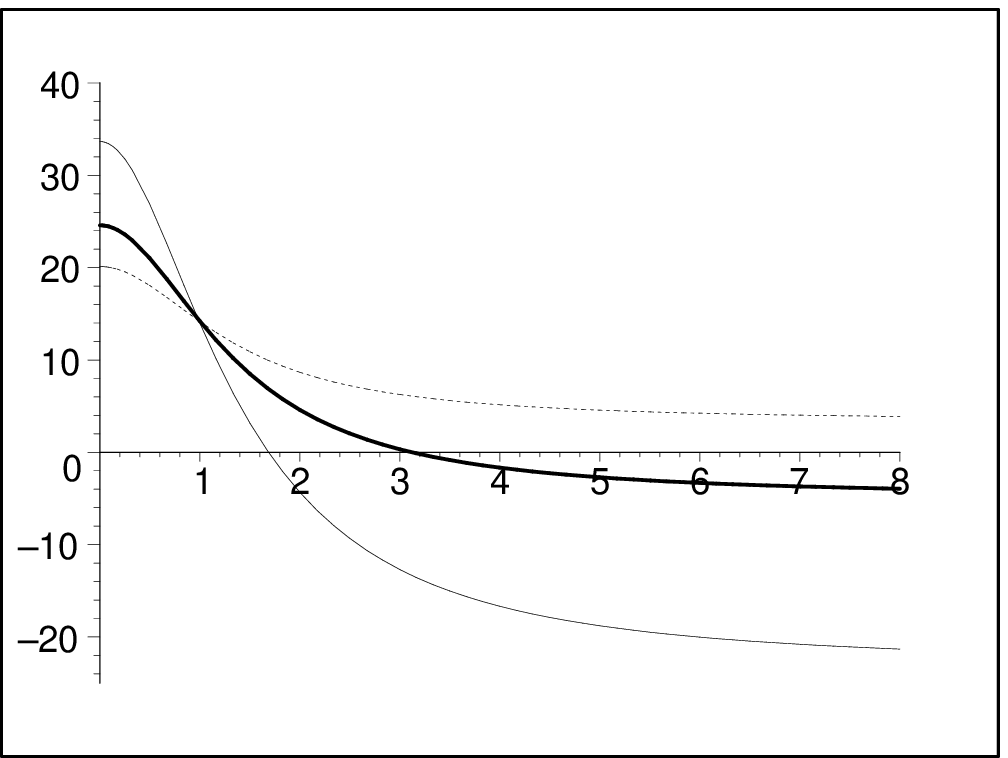} 
\includegraphics[scale=0.35]{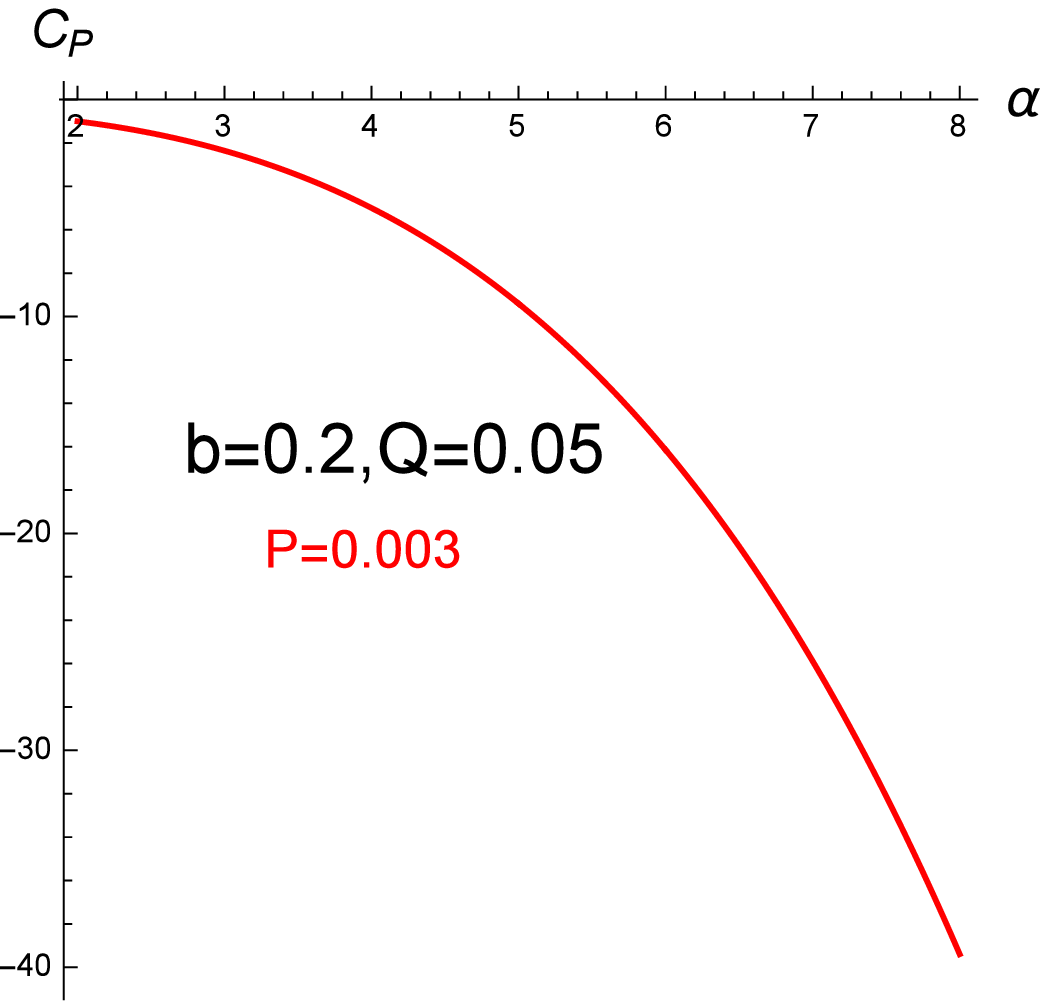}\\
\includegraphics[scale=0.35]{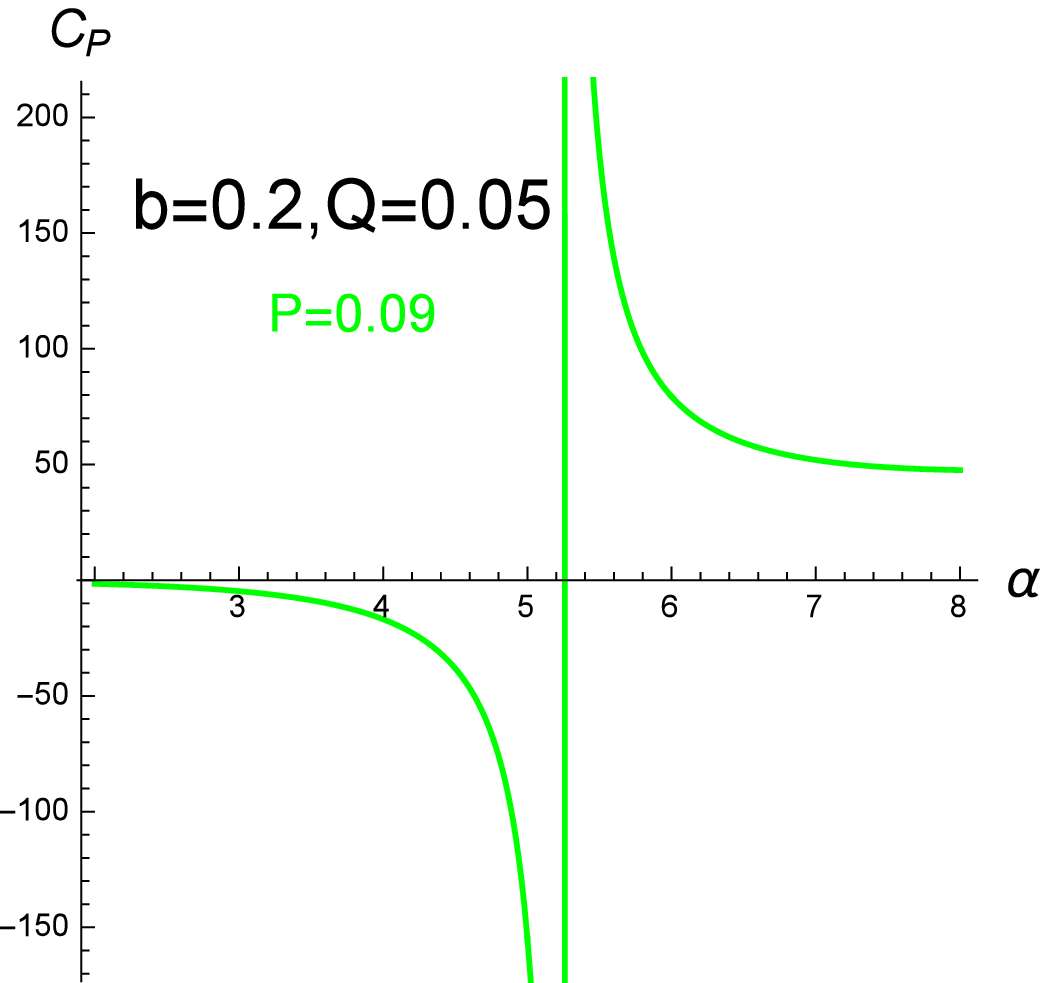}
\includegraphics[scale=0.35]{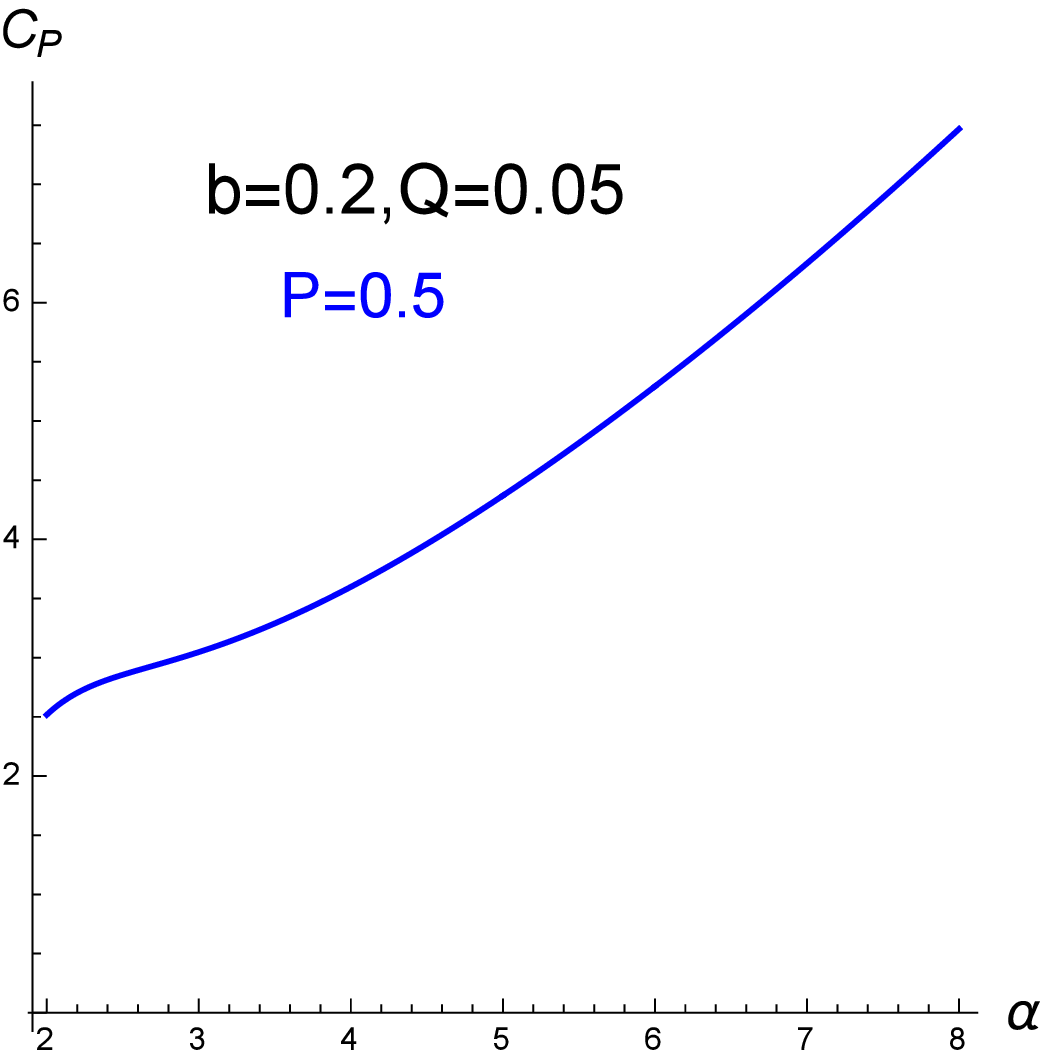}
\caption{(color online). The $\alpha$-related stability investigation in our work and ref.\cite{Sheykhi:2009pf}. The top-left part is copied from ref.\cite{Sheykhi:2009pf}, which displays the quantity $(\partial ^{2}M/\partial S^{2})_{Q}$ versus $\protect\alpha $ for $b=0.2$, $l=1$, $n=5$, $\rho_{+}=0.5$ (solid line), $\rho_{+}=0.55$ (bold line), and $\rho_{+}=0.6$ (dashed line). The other three sub-fig is our results for the behavior of $C_p$ versus $\protect\alpha $ with different pressures for $b=0.2$, $Q=0.05$ and $n=5$, $P=0.003$ (Red line), $P=0.09$ (green line), and $P=0.5$ (blue line) }
\label{CompareWithTheRef}
\end{center}
\end{figure}

Ref.\cite{Sheykhi:2009pf} analyzed the thermal stability of these black holes with the cosmological constant being fixed as a constant, whose results indicate that there exists a maximum value $\alpha _{max}$ for which the black hole solutions are thermadynamically unstable when $\alpha > \alpha _{max}$. However, our studies here,  especially the $C_p$-$T$ diagrams indicate that, for large $\alpha$'s, when the cosmological constant are considered dynamical degrees of freedom (namely pressures), stable phase is allowed in the appropriate region of parameters.  Ref.\cite{Sheykhi:2009pf} uses $(\partial ^{2}M/\partial S^{2})_{Q}$ as the criteria for thermodynamical stabilities, while we use the heat capacity $C_p$. Both this two quantities have the same functionality in analyzing  thermodynamical stabilities. But ref.\cite{Sheykhi:2009pf} does not fix the value of $Q$ in their $\frac{\partial^2M}{\partial S^2}|_Q$-$\alpha$ figure, see FIG.\ref{CompareWithTheRef} for comparisons. It fixex the value of $b,\rho _+, \Lambda$ and change the value of $\alpha$. From eq.$\eqref{Qnoc}$, we know that $Q=Q(b,\rho _+,\alpha,\Lambda)$. In this case, the value of $Q$ will change with $\alpha$. However, in our cases, we fix pressures $P$ and electric charges $Q$ according to eq.\eqref{CpabPQ} and show that stable phases (positive specific heat) are indeed possible when we change the value of pressures.

\section{QNMs or Dynamic Analysis \label{DynaAna}}  

In previous two sections, we explored variations of the phase structure of dilaton black holes as the coupling constant $\alpha$ varies. The results indicate that the Van der Waals like liquid-gas phase structure disappears from the system little by little when $\alpha$ exceeds a critical value, which we have denoted by $\alpha _1$. In gravitational pictures, it is the SBH phase that begins to disappear when $\alpha$ goes larger than $\alpha _1$, while the LBH phase continues to be stable no matter how large $\alpha$ grows into. According to the general idea of thermodynamic / blackhole corresbondence,  we expect that such variation trends should also be embodied in the change of QNFs either, refs \cite{Liu2014gvf,mahap1602,chab1606}. For example, ref.\cite{Liu2014gvf} uncovers that, differences between thermal-features of LBH and SBH phases have counter parts in QNFs. Just as we will show in the following, this fact can also be reproduced in the dialton black holes with some modulations by $\alpha$.

To see these parallelism or correspondence manifestly, we need to perturb the dilaton field $\Phi$ and calculate the QNFs numerically. Our calculation method is mainly the serial expansion of ref.\cite{horowitz1999}. However, it should be pointed out that the key features of this quasinormal modes in AdS black hole backgrounds is studied in the more earlier work \cite{adsQNM1997,adsQNM1999}. For convenience, we rewrite the metric \eqref{Metric} as follows
\bea{ReMetric}
&&\hspace{-7mm}ds^2=-F(\rho) dt^2 +\frac{d \rho ^2}{A(\rho) F(\rho)} + \rho ^2 H(\rho) d\Omega ^2 _{k,n-1}
\\
&&\hspace{-7mm}F(\rho) =N^2 (\rho) f^2 (\rho), A(\rho) =N^{-2} (\rho), H(\rho)=R^2 (\rho)
\eea
Equations of motion for the dilaton fields' perturbation in this background has the form, 
\bea{perofdilaton}
&\frac{\partial _\mu [\sqrt{-g} g^{\mu \nu} \partial _\nu (\delta \phi)] }{\sqrt{-g}} -\frac{n-1}{8} [ V''(\Phi_0) \delta \phi]
\nonumber\\ 
&-\frac{4\alpha ^2}{n-1} e^{-\frac{4\alpha \Phi _0(\rho)}{n-1}} F_{t \rho }^2 (\rho) g^{tt} g^{\rho \rho}\delta \phi=0
\eea
where $\Phi_0 (\rho)$ and $F_{t\rho} (\rho)$ are the classical solution of \eqref{phifield} and \eqref{solEM} respectively, and $V''(\Phi_0)$ is the second or derivative of \eqref{LP} respect to $\Phi$. We use symbols $\delta \phi$ to denote perturbations of $\Phi$ around the classical profile and write its general form as follows
\bea{decomposeOFdeltaPhi}    
&&\hspace{-5mm}\delta \phi (t,\rho,\vec{\theta}) = \sum _{\ell m}[\rho ^2 H(\rho)]^{\frac{2-d}{4}} \psi (\rho) Y_{\ell m}(\vec{\theta}) e^{-i \omega (t+\rho _*)}
\eea
where $\rho _*$ is defined as $d \rho _* = \frac{d \rho}{ \sqrt{A(\rho)} F(\rho)}$. This  ansatz for $\delta\phi$ has made the infalling condition of QNMs considered in the factor $e^{i\omega(t+\rho_*)}$ as long as $\psi(\rho)$ takes finite values on the horizon surface. Substituting \eqref{decomposeOFdeltaPhi} into $\eqref{perofdilaton}$, we will get radial equations for $\psi (\rho)$
\bea{}
&&A(\rho)F(\rho) \frac{d^2}{d \rho ^2} \psi(\rho) +[A(\rho) F'(\rho)+\frac{F(\rho) A'(\rho)}{2}
\label{QNMPerEqua}\\
&&-2i\omega \sqrt{A(\rho)}] \frac{d \psi (\rho)}{d\rho}-V (\rho) \psi (\rho)=0
\nonumber
\eea
\bea{}
&&\hspace{-5mm}V(\rho)=-\frac{4\pi P(1-\frac{b^2}{\rho^2})^{\frac{2\alpha ^2}{2+\alpha ^2}}}{3(b^2-\rho^2)^2 (2+\alpha ^2)^2} [-2b^4 \alpha ^2 (-4+\alpha ^2)      \\
\nonumber
&&\hspace{0mm}+b^2 \rho ^2(-6+\alpha ^2)(2+\alpha ^2)+\rho ^4(2+\alpha ^2)^2]\\
\nonumber
&&\hspace{0mm}-\frac{256 q^2 \alpha ^2 (1-\frac{b^2}{\rho ^2})^{\frac{4}{2+\alpha ^2}} A}{3 \pi ^2 \rho ^2 (b^2-\rho ^2)^2}\\
\nonumber
&&\hspace{0mm}-\frac{3AF[(H')^2-4H H'']}{16H^2}+\frac{3FH'[6A+\rho A']}{8 \rho H} \\
&&\hspace{0mm}+\frac{3AF'[2H+\rho H']}{4\rho H}+\frac{3F[A+\rho A']}{4\rho ^2} 
\nonumber
\eea
The standard definition of QNMs requires that the perturbation vanish at infinity as well as purely infalling in the near horizon region. For numeric conveniences, we change $\rho$ into $x=\frac{\rho _+}{\rho}$ and translate the integration region $\rho_+ < \rho < \infty$ into $ 0<x<1$. By the $x$-coordinate, eq.\eqref{QNMPerEqua} could be written as
\bea{emoincoorx}
[s(x) \psi''(x) +t(x) \psi'(x) +u(x) \psi (x)]=0 
\label{qnmEquation}
\eea{}
with coefficient functions $s(x)$, $t(x)$ and $u(x)$ defined correspondingly. The stratage of numeric QNMs is that, expand both the coefficient functions the goal function into serial form in the near horizon region
\begin{align}
\{s,t,u,\psi\}=\sum_{n=0}^\infty\{s_n,t_n,u_n,a_n\}(x-1)^n
\label{serialExpansion}
\end{align}
and assume that as long as higher enough order of expansions are taken, the value of goal function $\psi(x)$ following from the series would also be valid on the boundary surface thus satisfying conditions $\psi(0)=\sum a_n(-1)^n=0$. So the key task of finding QNMs is to substitute serial functions \eqref{serialExpansion} into \eqref{emoincoorx} and calculate the $\omega$-dependent coefficients $\{a_n^\omega\}$ to high enough order and zero points of their alternative summation $\sum a^\omega_n(-1)^n$. Using eqs.\eqref{serialExpansion} and \eqref{qnmEquation}, it can be easily verified that
\bea{}
\nonumber
&a_i = \sum ^{i-1} _{j=0} \frac{-1}{i[(i-1)s_1+t_0]} [~(i-j-2)(i-j-1) s_{j+2} \\
&\quad \quad  +(i-j-1)t_{j+1} +u_j~] a_{i-j-1}
\eea{} 
The leading coefficient $a_0$ is undetermined. But it does not affect the quasinormal frequencies $\omega$ due to linearities of the perturbation equation \eqref{emoincoorx}. With these preparations, we routinely calculate the QNFs numerically with fixed dilaton coupling $\alpha$ and the relevant thermodynamical quantaties $P,\rho _+(b,T,Q,\alpha),b(P,T,Q,\alpha)$.  

\begin{figure}[!ht]
\begin{center}
\includegraphics[scale=0.35]{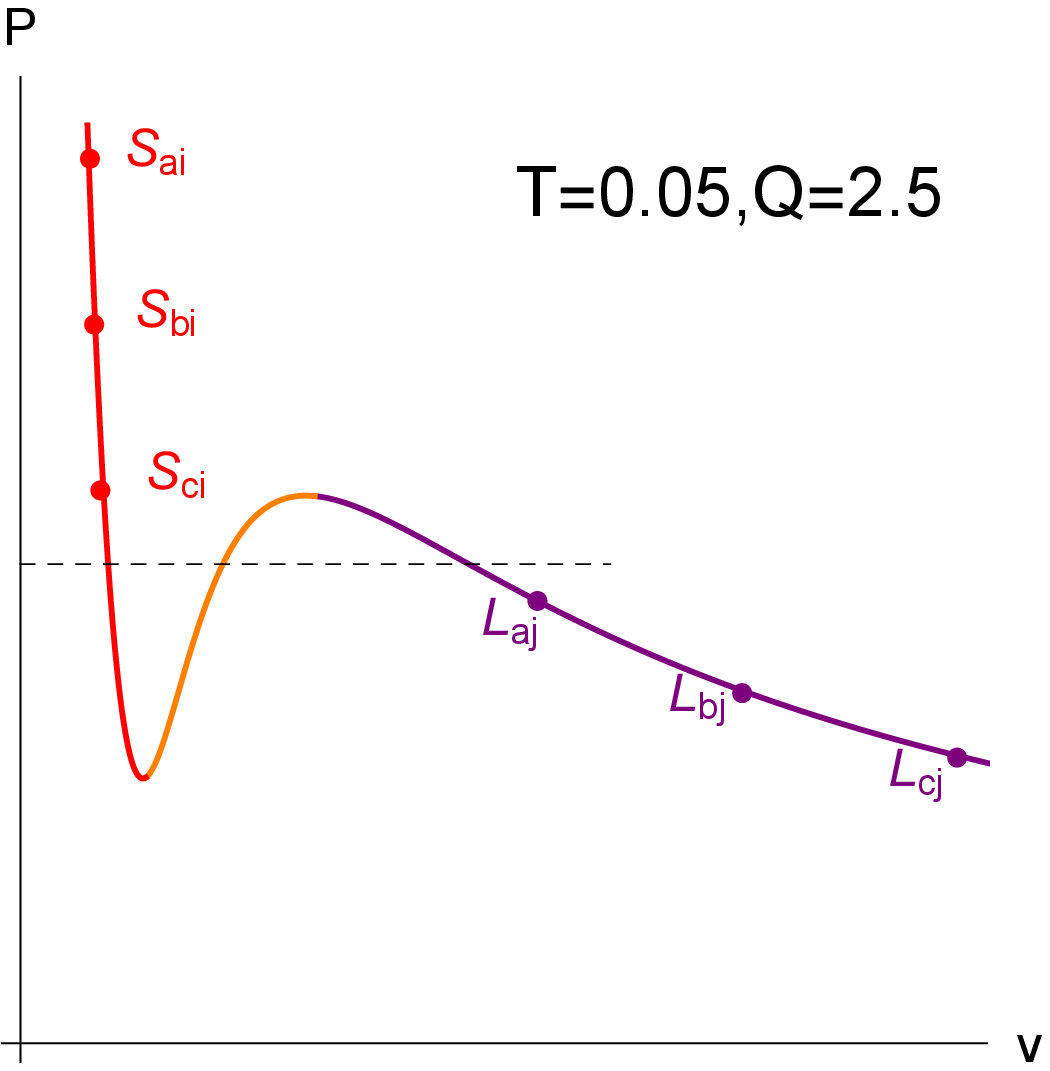}
\includegraphics[scale=0.35]{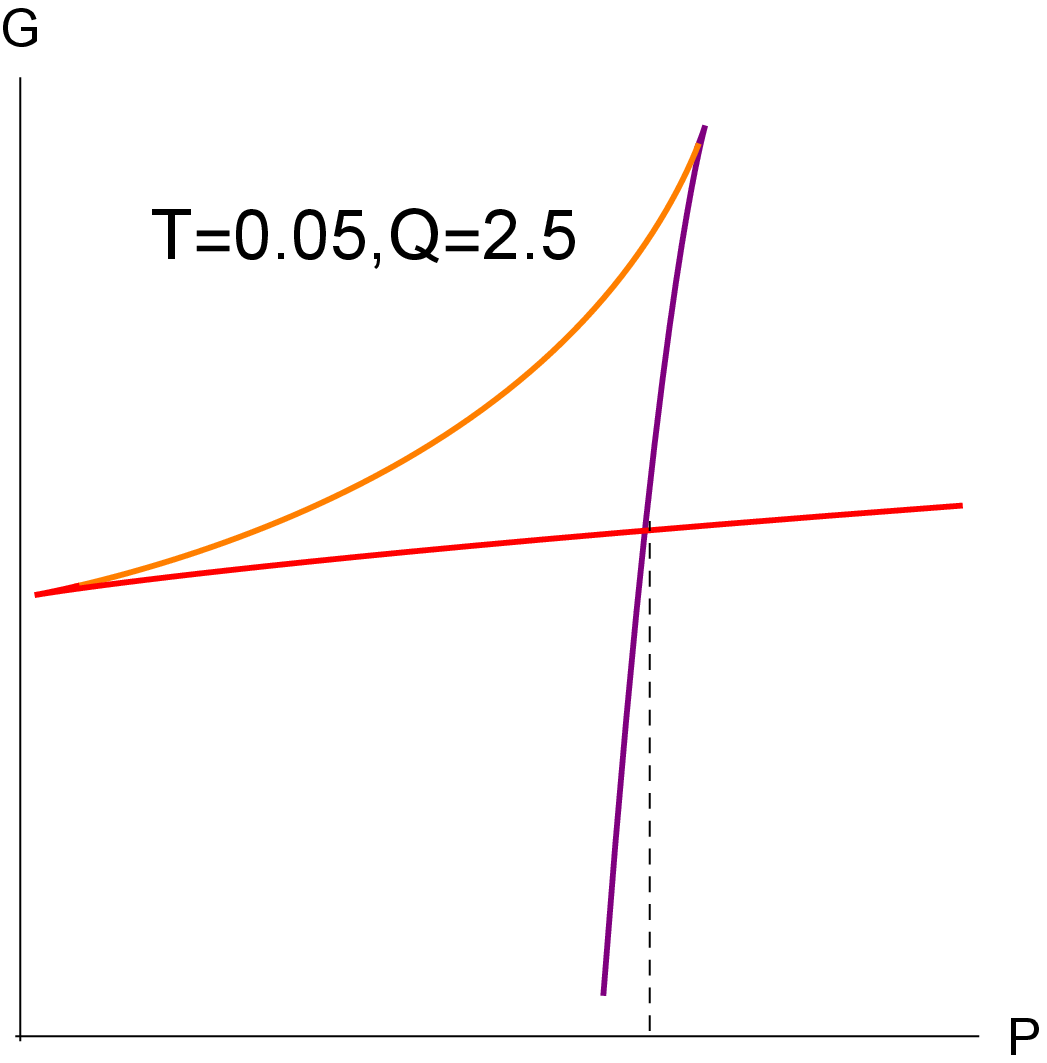}
\caption{The position of parameters in the $P$-v and $G$-$P$ line whose QNFs are presented in table. \ref{tablevsaLBHisocurve} and \ref{tablevsaSBHisocurve}.  In the LBH phase, $T=0.05,Q=2.5$ with $S$-subscript $a$, $b$, $c$ denoting three pressure values and $j=1,2,3,4,5$ five $\alpha$s $0.001,0.5,1,4/3,2$ respectively. In the SBH phase $T=0.05,Q=2.5$ with $L$-subscript $a$, $b$, $c$ denoting pressures and $j=1,2,3,4,5$ five $\alpha$s  $0.001,0.125,0.25,0.375,0.5$ respectively}
\label{comLBHinvaryalpha}
\end{center}
\end{figure} 
\begin{figure}[!ht]
\begin{center}
\includegraphics[scale=0.5]{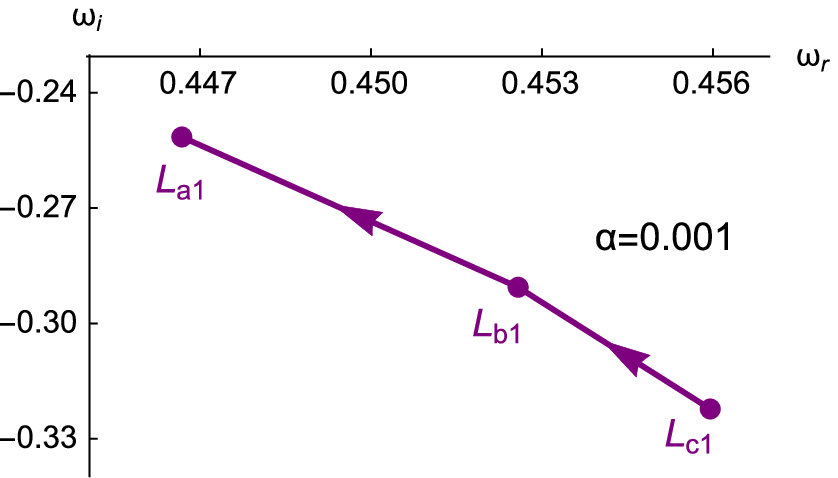}
\includegraphics[scale=0.5]{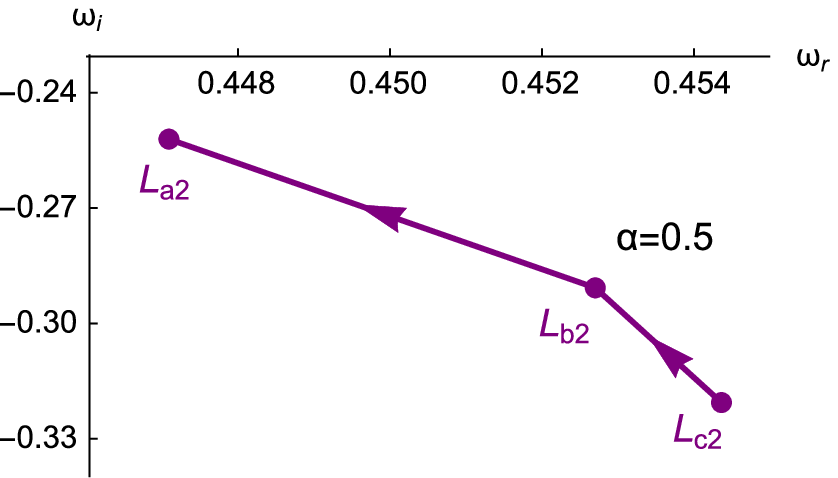}
\caption{QNFs as the pressure increases in the LBH phase, corresponding to $L_{c1} \to L_{b1} \to L_{a1}$ in Table.\ref{tablevsaLBHisocurve} with $\alpha$=0.001, the arrow indicates directions of the pressure's increasing. The right one corresponds to the case of $\alpha=0.5$}
\label{varyPLBH}
\end{center}
\end{figure} 
\begin{figure}[!ht]
\begin{center}
\includegraphics[scale=0.5]{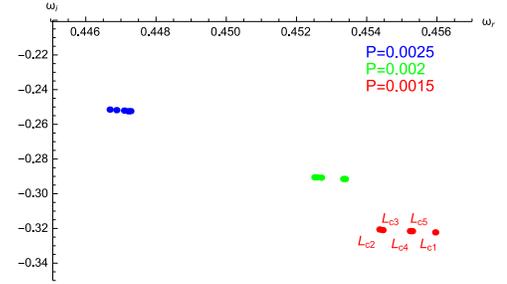}
\caption{ QNFs as the dilaton coupling $\alpha$ changes in LBH phase. Red, Green, Blue color part corresponds to $L_{c1} \to L_{c5}$, $L_{b1} \to L_{b5}$, $L_{a1} \to L_{a5}$ in Table.\ref{tablevsaLBHisocurve} with $P=0.0015, P=0.002, P=0.0025$ respectively.}
\label{varyaLBH}
\end{center}
\end{figure}

In table.\ref{tablevsaLBHisocurve} we list frequencies of the first \footnote{If we take 2nd or 3rd order QNMs, the results are similar.} order QNMs of dilaton perturbations in the LBH phase corresponding to isothermal lines in Fig.\ref{comLBHinvaryalpha} with T=0.05, Q=2.5.and $\alpha=0.001,0.5,1,1.33,2$ respectively.  It's easy to note that due to the negativeness of the quasinormal frequencies' imaginary part, referring FIG.\ref{varyPLBH} and \ref{varyaLBH}, such dilaton black holes are dynamically stable in the parameter region of LBH phases, see ref. \cite{konop0809,dfzeng0605} and related works. We also note that for LBH phases with equal $\alpha$s, as the horizon radius decreases or as the pressure increases along the isothermal line, both the real and absolute of the imaginary parts of the frequencies increase, referring FIG.\ref{varyPLBH}. While as $\alpha$ varies in this phase, the variation of the QNFs is almost negligible, referring FIG.\ref{varyaLBH}.
\begin{table}[!ht]
\begin{center}
\begin{tabular}{|c|c|c|c|c|}
		\hline
		    &$\alpha$&$P(10^{-3})$& $\rho _+ (b)$ & $\omega$ \\
		\hline
		$L_{a1}$&0.001&2.5&10.3964(b=0.242) & 0.4467-0.2515i  \\
		\hline
	    $L_{b1}$&0.001&2.0&14.6426(b=0.15)&0.4526-0.2906i \\
		\hline
		$L_{c1}$&0.001&1.5&21.2986(b=0.088)&0.4559-0.3223i\\
		\hline
		$L_{a2}$&0.5&2.5&10.4166(b=0.256) & 0.4471-0.2521i  \\
		\hline
		$L_{b2}$&0.5&2.0&14.6492(b=0.159)&0.4527-0.2907i \\
		\hline
		$L_{c2}$&0.5&1.5&21.194(b=0.094)&0.4544-0.3206i\\
		\hline
		$L_{a3}$&1&2.5&10.4072(b=0.296) & 0.44687-0.25182i  \\
		\hline
		$L_{b3}$&1&2.0&14.6833(b=0.183)&0.4533-0.2915i \\
		\hline
		$L_{c3}$&1&1.5&21.1998(b=0.1085)&0.4545-0.3207i\\
		\hline
		$L_{a4}$&4/3&2.5&10.4275(0.3313)&0.44727-0.2524i   \\
		\hline
		$L_{b4}$&4/3&2.0&14.6863(0.2053)&0.45338-0.29155i \\
		\hline
		$L_{c4}$&4/3&1.5&21.2552(0.1213)&0.45529-0.32158i\\
		\hline
		$L_{a5}$&2&2.5&10.4281(0.4176)&0.4472-0.25247i   \\
		\hline
		$L_{b5}$&2&2.0&14.6401(0.2599)&0.45251-0.29057i \\
		\hline
		$L_{c5}$&2&1.5&21.2514(0.15291)&0.45523-0.32152i\\
		\hline
\end{tabular}
\caption{This table shows how quasinormal frequencies change as the black hole horizon size $\rho _+$ increases in the LBH phase along the isothermal line with $T=0.05,Q=2.5$ with three different $\alpha$s $0.001,0.5,1,4/3,2$}
\label{tablevsaLBHisocurve}
\end{center}
\end{table}

\begin{figure}[!ht]
\begin{center}
\includegraphics[scale=0.5]{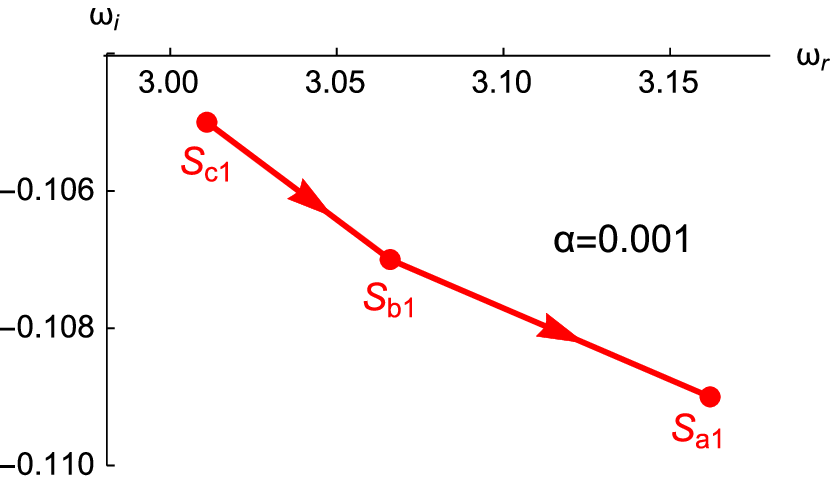}
\includegraphics[scale=0.5]{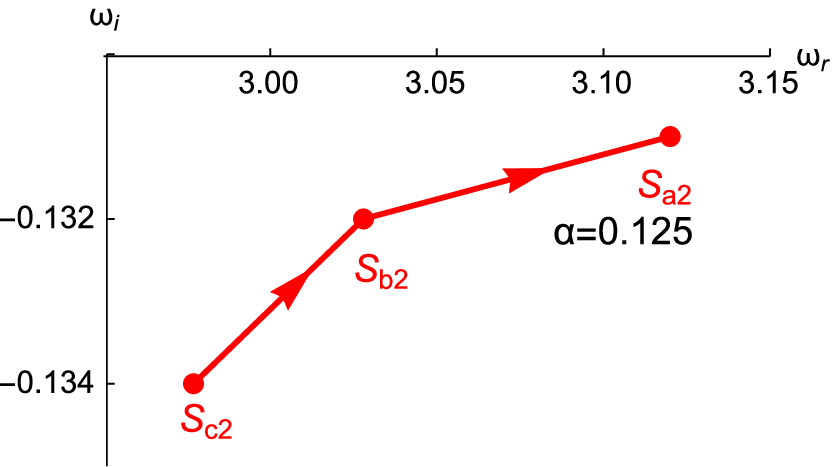}
\caption{QNFs as the pressure increases in the SBH phase, corresponding to $S_{c1} \to S_{b1} \to S_{a1}$ in Table.\ref{tablevsaSBHisocurve} with $\alpha$=0.001, the arrow indicates directions of the pressure increasing. The right one corresponds to cases with $\alpha=0.125$}
\label{varySBH}
\end{center}
\end{figure}
\begin{figure}[!ht]
\begin{center}
\includegraphics[scale=0.5]{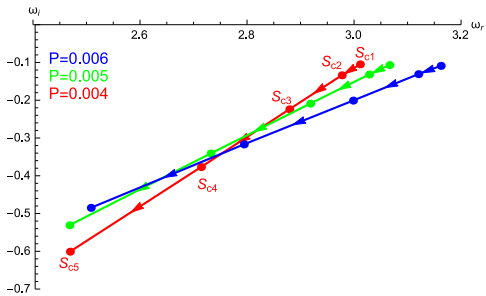}
\caption{QNFs as the dilaton coupling $\alpha$ changes in the SBH phase. Red, green, blue points corresponds to dataset $S_{c1} \to S_{c5}$, $S_{b1} \to S_{b5}$, $L_{a1} \to L_{a5}$  in table.\ref{tablevsaSBHisocurve}  with $P=0.004, P=0.005, P=0.006$ respectively. Arrow is the direction of $\alpha$'s increasing.}
\label{QNMfreSBH}
\end{center}
\end{figure}

In table. \ref{tablevsaSBHisocurve} we list the third \footnote{The reason we take 3rd order quasinormal modes in SBH phases is that, as $\alpha$ increases, the $1$st and $2$nd order QNM's frequency quickly flows to the $\omega_r=0$ line.} order QNFs for some typical small black holes along the isothermal line with T=0.05 and $\alpha=0.001$, $0.125$, $0.25$, $0.375$, $0.5$ respectively. Similiar to the LBH phase, we note that all such dilaton black holes are dynamically stable in the parameter region we choose due to the negativeness of the quasinormal frequencies' imagary part, FIG.\ref{varySBH} and \ref{QNMfreSBH}. 
\begin{table}[!ht]
	\begin{center}
		\begin{tabular}{|c|c|c|c|c|}
			\hline
			   &$\alpha$&$P(10^{-3})$& $\rho _+(b)$ & $\omega$ \\
			\hline
			$S_{a1}$&0.001&6&2.22632(1.491)&3.162-0.109i \\
			\hline
			$S_{b1}$&0.001&5&2.27789(1.4803)&3.066-0.107i   \\
			\hline
			$S_{c1}$&0.001&4&2.34339(1.4621)&3.011-0.105i   \\
			\hline
			$S_{a2}$&0.125&6&2.22675(1.4974)&3.12-0.131i   \\
			\hline
			$S_{b2}$&0.125&5&2.2779(1.4867)&3.028-0.132i    \\
			\hline
			$S_{c2}$&0.125&4&2.34302(1.4685)&2.977-0.134i \\
			\hline			
			$S_{a3}$&0.25&6&2.22859(1.5164)&2.998-0.201i  \\
			\hline
			$S_{b3}$&0.25&5&2.27916(1.5055)&2.918-0.209i  \\
			\hline
			$S_{c3}$&0.25&4&2.3432(1.4872)&2.879-0.224i \\
			\hline
			$S_{a4}$&0.375&6&2.23176(1.5477)&2.794-0.317i \\
			\hline
			$S_{b4}$&0.375&5&2.280829(1.5366)&2.732-0.341i \\
			\hline
			$S_{c4}$&0.375&4&2.34361(1.518)&2.714-0.377i \\
			\hline
			$S_{a5}$&0.5&6&2.23589(1.5909)&2.508-0.485i  \\
			\hline
			$S_{b5}$&0.5&5&2.28307(1.5795)&2.468-0.531i \\
			\hline
			$S_{c5}$&0.5&4&2.34353(1.5607)&2.469-0.601i\\
			\hline
\end{tabular}
\caption{This table shows how the quasinormal frequencies change as the black hole horizons $\rho _+$ increase for small black holes phase in the isothermal curves with $T=0.05,Q=2.5$ under the $\alpha=0.001,0.125,0.25,0.375,0.5$ respectively}
\label{tablevsaSBHisocurve}
\end{center}
\end{table}

\begin{figure}[!ht]
\begin{center}
\includegraphics[scale=0.45]{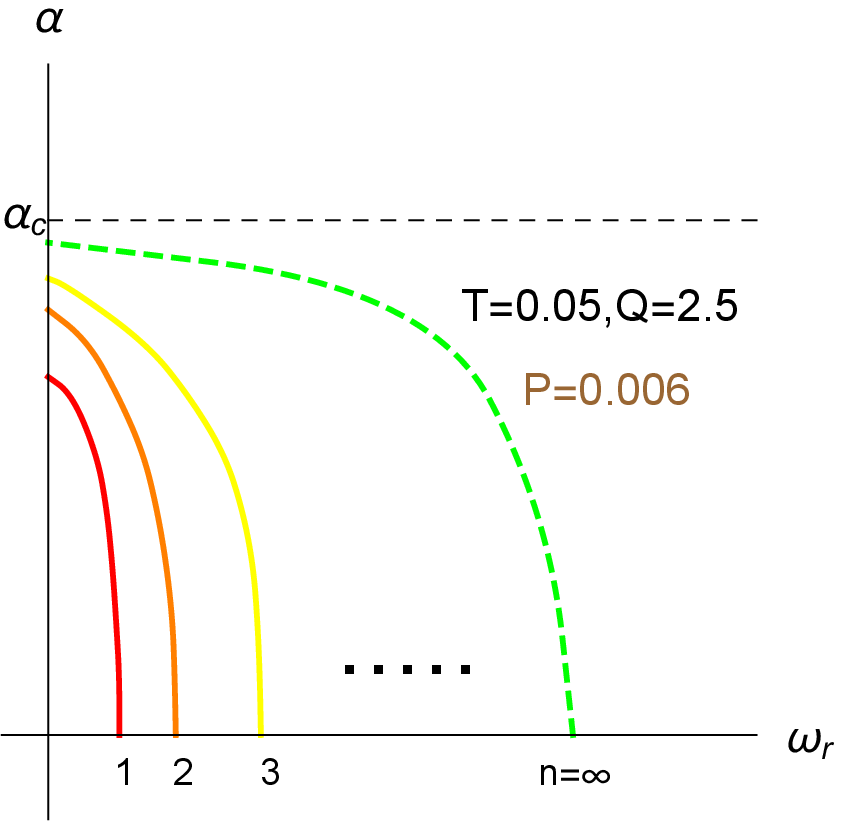}
\includegraphics[scale=0.45]{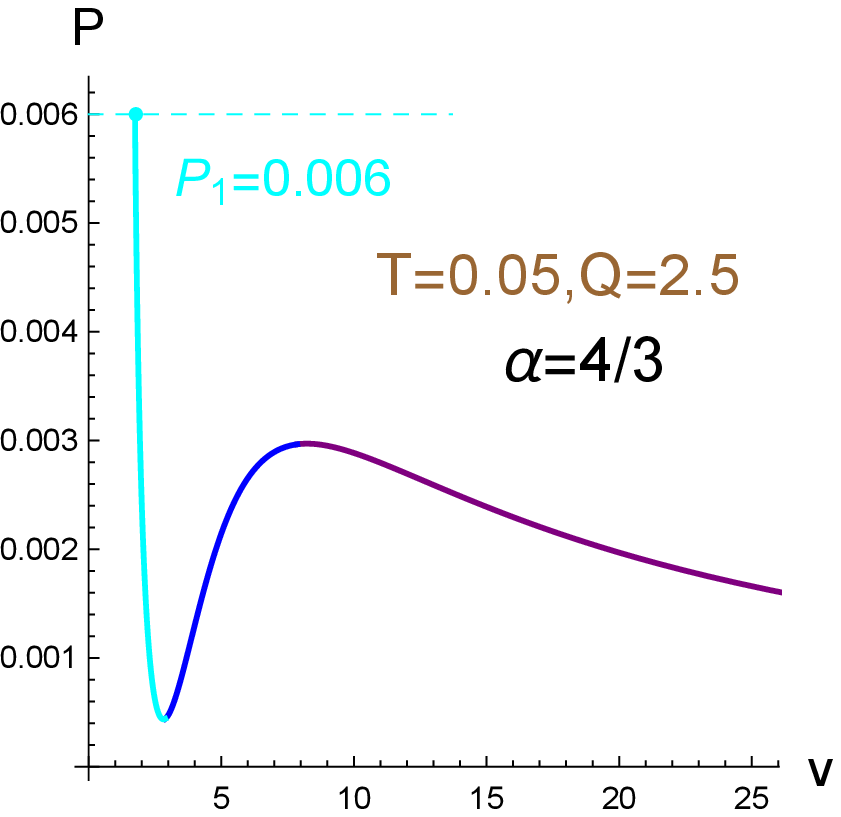}
\caption{The left panel is the evolving trends of 1st, 2nd and 3rd order QNFs' real part in the SBH phase featured by thermal parameters $P_1=0.006$, $T=0.05$ and $Q=2.5$. $\alpha _c\approx1.33$ is an upper limit of $\alpha$ approaching which the real part of all QNFs flows to zero. The right panel shows isothermo line of the system featured by parameters $T=0.05,Q=2.5$ and $\alpha =4/3$. Pressures in this line is bounded from above by $P<P_1=0.006$.}
\label{SBHac}
\end{center}
\end{figure}

Comparing Fig.\ref{varyPLBH} and \ref{varySBH}, we observe that similar phenomena occurs to the dilaton black hole as well simple ones of ref.\cite{Liu2014gvf}. That is, as two thermodynamically stable phase in the Van der Waals like system, SBH and LBH phase indeed have different evolving trends in their QNFs as one goes along the isothermal line towards pressure increasing direction. In the LBH phase, the QNFs decrease as the pressure increases. While in the SBH phase, the trends just reverses. Of course, the quantitative trends in the dilaton black holes are $\alpha$ dependent. 

In the previous section, we observed that the LBH and SBH phase has different evolving trends as $\alpha$ becomes large. That is, the LBH continues to exist and be stable, but the SBH experience a process of disappearing from the system. From FIG.\ref{varyaLBH} and \ref{QNMfreSBH}, we see that this phenomena also has counter parts in the QNFs. That is, in the LBH phase, the QNF is almost $\alpha$-independent while in the SBH phase, the case becomes remarkably different. We tracked in the left part of FIG.\ref{SBHac} evolutions of the real part of 1st, 2nd and 3rd order QNFs in the SBH phase as $\alpha$ varies, from which we see that as $\alpha$ goes to some upper limit $\alpha_c$, the real part of these QNFs all flows to zero uniformly. 

\section{Conclusions and Discussion}

We investigate in this work phase structures of a charged AdS dilaton black hole constructed in \cite{Gao:2004tu,Gao:2004tv,Gao:2005xv} from both thermal ensemble and dynamical aspects and get results consistent to each other. In thermal ensemble sides, we treat the cosmological constant as independent dynamic variables and derive out it's conjugate partner - thermal volumes and check the generalized first law of black hole thermodynamics in the extended phase space. The dilaton coupling $\alpha$ measures the strength of string corrections to the Einstein-Maxwell gravitations. To see this correction effects, we compare the $P$-v, $G$-$P$ and $C_p$-$T$ lines for some typical values $\alpha \ne 0$ (corrected gravitation theories) with those of $\alpha =0$ (Einstein-Maxwell gravity, van der Waals like phase structure) in canonical thermal ensembles. In small $\alpha$ cases, phase structures in the dilaton theory are smoothly joined with those of Einstein-Maxwell theory, comparing FIG.\ref{PVwithalpha0}, \ref{CpTwithalpha0fixP1toP3}, \ref{GPwithalpha0} with the relevant figures from \cite{Kubiznak1205} and ref.\cite{fernado1611}. However, as $\alpha$ increases, new phase structures appear in the $P$-v plane, FIG.\ref{PVwithalpha2}, which is characterized by the lacking of middle sized black hole configuration. As $\alpha$ increases further, the Van der Waals like phase structure featured by 3-size BH's existence disappears, only the new phase structure featured by the 2-size BHs' existence is left, FIG.\ref{Pvwithalpha8} and \ref{CpTwithalpha8}. Basing on these results, we conclude that the dilaton or string effect on the phase structure of BHs in the corresponding gravitation theory is remarkable.

In dynamic considerations, we perturb the dilaton field $\Phi$ and study their QNMs on series of BHs along the isothermal line with varyious dilaton couplings $\alpha$. Fixing $\alpha$, we compare evolving trends of SBH and LBH's QNFs, and see totally different features between the two, Fig.\ref{varyPLBH} and Fig.\ref{varySBH}. This tells us that, as two thermodynamic phase, they also display different features under dynamic perturbations. This is consistent with results reported in the literature \cite{Liu2014gvf}. But in our cases, modulated by the dilaton coupling constant $\alpha$. Furthermore, we observe new parallelism between the feature of QNMs and thermal phases, FIG.\ref{SBHac}.  In thermal ensemble sides, as $\alpha$ increases, the SBHs' Van der Waals like isothermal line experience a process of losing the higher pressure part, while the LBHs do not. In dynamic considerations, the ``laziness'' of LBHs manifests as the almost $\alpha-$independence of QNFs, Fig.\ref{varyaLBH}. While the SBHs' QNF manifests very sensitive dependence on $\alpha$, Fig.\ref{QNMfreSBH}. This parallelism between thermal phase structure and dynamic features of the spacetime geometry can be seen as new words of the more basic principle of thermodynamic/blackholes or Gauge/Gravity duality.

There is statements in the literature that the charged dilaton black hole is thermodynamically unstable in the large $\alpha$ case basing on analysis of the $\partial ^2 M/\partial S^2$ versus $\alpha$ relation with fixed value of cosmological constant. However, our studies basing on both thermal ensemble and dynamic perturbation analysis indicate that, when the cosmological constant are treated as independent thermodynamic variables, this instability will disappear in appropriate parameter ranges.  Supports from the former can be seen from the positiveness of capacities of FIG.\ref{CompareWithTheRef}, while supports from the latter can be seen from the negativeness of the QNFs in various perturbations in FIG.\ref{varyPLBH},\ref{varyaLBH},\ref{varySBH},\ref{QNMfreSBH}.

As discussions, we point out here that the following extensions are still worthwhile to explore. The first is, thermodynamics of other charged AdS dilaton black holes in the extend phase space methods and whether or not our results depend on the model of gravitation theories sensitively. The second is, dynamical analysis of more general perturbation on both the dilaton and metric fields. Such analysis may reveal more comprehensively stability issues of these charged dilaton black holes. Finally, to find whether or not true thermal systems dual to these charged dilaton black holes exist in nature is also a valuable question.


\begin{thebibliography}{99}

\bibitem{Maldacena:1997re} 
J.~M.~Maldacena,
``The Large N limit of superconformal field theories and supergravity,''
{\it Int. J. Theor. Phys.} {\bf 38}, 1113 (1999)
{\it Adv. Theor. Math. Phys.} {\bf 2}, 231 (1998),
\href{https://arxiv.org/abs/hep-th/9711200}{arXiv: hep-th/9711200}

\bibitem{Witten:1998qj} 
E.~Witten,
``Anti-de Sitter space and holography,''
{\it Adv. Theor. Math. Phys.} {\bf 2}, 253 (1998),
\href{https://arxiv.org/abs/hep-th/9802150}{arXiv: hep-th/9802150}.

\bibitem{Hawking:1982dh} 
S.~W.~Hawking and D.~N.~Page,
``Thermodynamics of Black Holes in anti-De Sitter Space,''
{\it Commun. Math. Phys.} {\bf 87}, 577 (1983).

\bibitem{Cai:1998ji} 
R.~G.~Cai and K.~S.~Soh,
``Critical behavior in the rotating D-branes,''
{\it Mod. Phys. Lett.A} {\bf 14}, 1895 (1999),
\href{https://arxiv.org/abs/hep-th/9812121}{arXiv: hep-th/9812121}

\bibitem{Cvetic:1999ne}   
M.~Cvetic and S.~S.~Gubser,
``Phases of R charged black holes, spinning branes and strongly coupled gauge theories,''
{\it JHEP.} {\bf 9904}, 024 (1999),
\href{https://arxiv.org/abs/hep-th/9902195}{arXiv: hep-th/9902195}

\bibitem{Cvetic:2001bk} 
M.~Cvetic, S.~Nojiri and S.~D.~Odintsov,
``Black hole thermodynamics and negative entropy in de Sitter and anti-de Sitter Einstein-Gauss-Bonnet gravity,''
{\it Nucl. Phys. B} {\bf 628}, 295 (2002),
\href{https://arxiv.org/abs/hep-th/0112045}{arXiv: hep-th/0112045}

\bibitem{Cai:2001dz} 
R.~G.~Cai,
``Gauss-Bonnet black holes in AdS spaces,''
{\it Phys. Rev. D} {\bf 65}, 084014 (2002),
\href{https://arxiv.org/abs/hep-th/0109133}{arXiv: hep-th/0109133}

\bibitem{Gibbons:1987ps} 
G.~W.~Gibbons and K.~i.~Maeda,
``Black Holes and Membranes in Higher Dimensional Theories with Dilaton Fields,''
{\it Nucl. Phys. B} {\bf 298}, 741 (1988).

\bibitem{Garfinkle:1990qj} 
D.~Garfinkle, G.~T.~Horowitz and A.~Strominger,
``Charged black holes in string theory,''
{\it Phys. Rev. D }{\bf 43}, 3140 (1991),
Erratum: [Phys. Rev. D {\bf 45}, 3888 (1992)].

\bibitem{Gregory:1992kr}       
R.~Gregory and J.~A.~Harvey,
``Black holes with a massive dilaton,''
{\it Phys. Rev. D} {\bf 47}, 2411 (1993),
\href{https://arxiv.org/abs/hep-th/9210012}{arXiv: hep-th/9209070}

\bibitem{Horne:1992bi} 
J.~H.~Horne and G.~T.~Horowitz,
``Black holes coupled to a massive dilaton,''
{\it Nucl. Phys. B} {\bf 399}, 169 (1993),
\href{https://arxiv.org/abs/hep-th/9210012}{arXiv: hep-th/9210012}

\bibitem{Poletti:1994ff} 
S.~J.~Poletti and D.~L.~Wiltshire,
``The Global properties of static spherically symmetric charged dilaton space-times with a Liouville potential,''
{\it Phys. Rev. D} {\bf 50}, 7260 (1994),
Erratum: [Phys. Rev. D {\bf 52}, 3753 (1995)],
\href{https://arxiv.org/abs/gr-qc/9407021}{arXiv: gr-qc/9407021}

\bibitem{Mignemi:1991wa} 
S.~Mignemi and D.~L.~Wiltshire,
``Black holes in higher derivative gravity theories,''
{\it Phys. Rev. D} {\bf 46}, 1475 (1992),
\href{https://arxiv.org/abs/hep-th/9202031}{arXiv: hep-th/9202031}

\bibitem{Gao:2004tu} 
C.~J.~Gao and S.~N.~Zhang,
``Dilaton black holes in de Sitter or Anti-de Sitter universe,''
{\it Phys. Rev. D} {\bf 70}, 124019 (2004),
\href{https://arxiv.org/abs/hep-th/0411104}{arXiv: hep-th/0411104}

\bibitem{Gao:2004tv}  
C.~J.~Gao and S.~N.~Zhang,
``Higher dimensional dilaton black holes with cosmological constant,''
{\it Phys. Lett. B} {\bf 605}, 185 (2005),
\href{https://arxiv.org/abs/hep-th/0411105}{arXiv: hep-th/0411105}

\bibitem{Gao:2005xv} 
C.~J.~Gao and S.~N.~Zhang,
``Topological black holes in dilaton gravity theory,''
{\it Phys. Lett. B }{\bf 612}, 127 (2005).

\bibitem{Gao:2006fu} 
C.~J.~Gao and S.~N.~Zhang,
``A Universe Dominated by Dilaton Field,''  
\href{https://arxiv.org/abs/astro-ph/0605682}{arXiv: astro-ph/0605682}

\bibitem{Zhang:2015dia} 
S.~J.~Zhang and E.~Abdalla,
``Holographic Thermalization in Charged Dilaton Anti-de Sitter Spacetime,''
{\it Nucl. Phys. B} {\bf 896}, 569 (2015),
\href{https://arxiv.org/abs/1503.07700}{arXiv:1503.07700 [hep-th]}

\bibitem{Sheykhi:2009pf} 
A.~Sheykhi, M.~H.~Dehghani and S.~H.~Hendi,
``Thermodynamic instability of charged dilaton black holes in AdS spaces,''
{\it Phys. Rev. D} {\bf 81}, 084040 (2010),
\href{https://arxiv.org/abs/0912.4199}{arXiv: 0912.4199 [hep-th]}

\bibitem{Sheykhi:2016syb} 
A.~Sheykhi, S.~H.~Hendi, F.~Naeimipour, S.~Panahiyan and B.~Eslam Panah,
``Thermodynamic geometry of charged dilaton black holes in AdS spaces,''
{\it Can. J. Phys.} {\bf 94}, no. 10, 1045 (2016).

\bibitem{adsLambdaThermo1}    
D. Kastor, S. Ray, J. Traschen,
``Enthalpy and the Mechanics of AdS Black Holes'',
{\em Class.Quant.Grav.} {\bf26}, 195011 (2009),
\href{https://arxiv.org/abs/0904.2765}{arXiv:0904.2765}.

\bibitem{Kubiznak1205} 
D.~Kubiznak and R.~B.~Mann,
``P-V criticality of charged AdS black holes'',
{\it JHEP} {\bf 1207}, 033 (2012),
\href{https://arxiv.org/abs/1205.0559}{arXiv:1205.0559 [hep-th]}

\bibitem{ccly1306} 
R.~G.~Cai, L.~M.~Cao, L.~Li and R.~Q.~Yang,
``P-V criticality in the extended phase space of Gauss-Bonnet black holes in AdS space,''
{\it JHEP} {\bf 1309}, 005 (2013),
\href{https://arxiv.org/abs/1306.6233}{arXiv:1306.6233}.

\bibitem{panahi1506}
S. H. Hendi, S. Panahiyan, B. Eslam Panah, M. Momennia,
"Phase transition of charged black holes in massive gravity through new methods",
{\em Ann. Phys.} {\bf528} (2016) {819}
\href{https://arxiv.org/abs/1506.07262}{arXiv: 1506.07262}

\bibitem{panahi1510}
S. H. Hendi, B. Eslam Panah, S. Panahiyan,
"Thermodynamical Structure of AdS Black Holes in Massive Gravity with Stringy Gauge-Gravity Corrections",
{\em Class. Quantum Grav.} {\bf33} (2016) 235007,
\href{https://arxiv.org/abs/1510.00108}{arXiv: 1510.00108}

\bibitem{panahi1511}
S. H. Hendi, S. Panahiyan, B. Eslam Panah,
"Extended phase space of Black Holes in Lovelock gravity with nonlinear electrodynamics",
{\em Prog. Theor. Exp. Phys.} {\bf2015} (2015) 103E01
\href{https://arxiv.org/abs/1511.00656}{arXiv: 1511.00656}

\bibitem{Kubiznak1608}  
D.~Kubiznak, R.~B.~Mann and M.~Teo,
``Black hole chemistry: thermodynamics with Lambda,''
{\it Class.Quant. Grav.}  {\bf 34}, no. 6, 063001 (2017),
\href{https://arxiv.org/abs/1608.06147}{arXiv: 1608.06147}.

\bibitem{panahi1608}
S. H. Hendi, G.Q. Li, J.X. Mo, S. Panahiyan, B. E. Panah,
"New perspective for black hole thermodynamics in Gauss-Bonnet-Born-Infeld massive gravity",
{\em Eur. Phys. J.} {\bf C76} (2016) 571
\href{https://arxiv.org/abs/1608.03148}{arXiv: 1608.03148}

\bibitem{fernado1611} 
S.~Fernando,
``P-V criticality in AdS black holes of massive gravity,''
{\it Phys. Rev. D} {\bf 94}, no. 12, 124049 (2016),
\href{https://arxiv.org/abs/1611.05329}{arXiv:1611.05329}

\bibitem{panahi1702}
S. H. Hendi, R. B. Mann, S. Panahiyan, B. Eslam Panah,
"van der Waals like behaviour of topological AdS black holes in massive gravity",
{\em Phys. Rev.} {\bf D95} (2017) 021501(R),
\href{https://arxiv.org/abs/1702.00432}{arXiv:1702.00432}.

\bibitem{Altamirano:2013uqa} 
N.~Altamirano, D.~KubizÅÃ¡k, R.~B.~Mann and Z.~Sherkatghanad,
``Kerr-AdS analogue of triple point and solid/liquid/gas phase transition,''
{\it Class. Quant. Grav.}  {\bf 31}, 042001 (2014),
\href{https://arxiv.org/abs/1308.2672}{arXiv:1308.2672}

\bibitem{Gunasekaran:2012dq}  
S.~Gunasekaran, R.~B.~Mann and D.~Kubiznak,
``Extended phase space thermodynamics for charged and rotating black holes and Born-Infeld vacuum polarization,''
{\it JHEP }{\bf 1211}, 110 (2012),
\href{https://arxiv.org/abs/1208.6251}{arXiv:1208.6251}

\bibitem{zzmz1305} 
R.~Zhao, H.~H.~Zhao, M.~S.~Ma and L.~C.~Zhang,
``On the critical phenomena and thermodynamics of charged topological dilaton AdS black holes,''
{\it Eur. Phys. J. C }{\bf 73}, 2645 (2013),
\href{https://arxiv.org/abs/1305.3725}{arXiv:1305.3725}.

\bibitem{zzmz1403} 
H.~H.~Zhao, L.~C.~Zhang, M.~S.~Ma and R.~Zhao,
``P-V criticality of higher dimensional charged topological dilaton de Sitter black holes,''
{\it Phys. Rev. D} {\bf 90} (2014) 064018.

\bibitem{panahi1503}
S. H. Hendi, G. H. Bordbar, B. Eslam Panah, M. Najafi,
"Dilatonic Equation of Hydrostatic Equilibrium and Neutron Star Structure"
{\em Astrophys Space Sci.} {\bf 358} (2015) 30,
\href{https://arxiv.org/abs/1503.01011}{arXiv: 1503.01011}.

\bibitem{panahi1509}
S. H. Hendi, A. Sheykhi, S. Panahiyan, B. Eslam Panah,
"Phase transition and thermodynamic geometry of Einstein-Maxwell-dilaton black holes"
{\em Phys. Rev.} {\bf D92} (2015) 064028,
\href{https://arxiv.org/abs/1509.08593}{arXiv:1509.08593}.

\bibitem{mlx1601}
J.~X.~Mo, G.~Q.~Li and X.~B.~Xu,
``Effects of power-law Maxwell field on the critical phenomena of higher dimensional dilaton black holes,''
{\it Phys. Rev. }{\bf D93} (2016) 084041,
\href{https://arxiv.org/abs/1601.05500}{arXiv: 1601.05500}

\bibitem{panahi1609}
"Three dimensional dilatonic gravity's rainbow: exact solutions",
S. H. Hendi, B. Eslam Panah, S. Panahiyan
{\em PTEP} {\bf2016} (2016) 103A02,
\href{https://arxiv.org/abs/1609.02002}{arXiv: 1609.02002}.

\bibitem{panahi1703}
S. H. Hendi, B. Eslam Panah, S. Panahiyan, A. Sheykhi,
"Dilatonic BTZ black holes with power-law field"
{\em Phys. Lett.} {\bf B767} (2017) 214,
\href{https://arxiv.org/abs/1703.03403}{arXiv: 1703.03403}.

\bibitem{Liu2014gvf} 
Y.~Liu, D.~C.~Zou and B.~Wang,
``Signature of the Van der Waals like small-large charged AdS black hole phase transition in quasinormal modes,''
{\it JHEP} {\bf 1409}, 179 (2014)
\href{https://arxiv.org/abs/1405.2644}{arXiv:1405.2644}.

\bibitem{mahap1602}
S. Mahapatra,
``Thermodynamics, Phase Transition and Quasinormal modes with Weyl corrections'',
{\em JHEP} {\bf04}, 142 (2016),
\href{https://arxiv.org/abs/1602.03007}{1602.03007}.

\bibitem{chab1606} 
M.~Chabab, H.~El Moumni, S.~Iraoui and K.~Masmar,
``Behavior of quasinormal modes and high dimension RNâAdS black hole phase transition,''
{\it Eur.\ Phys.\ J.\ C}{\bf 76}, no. 12, 676 (2016)
\href{https://arxiv.org/abs/1606.08524}{arXiv:1606.08524}.

\bibitem{Dayyani:2016gaa} 
Z.~Dayyani, A.~Sheykhi and M.~H.~Dehghani,
``Counterterm method in Einstein dilaton gravity and the critical behavior of dilaton black holes with a power-Maxwell field,''
{\it Phys. Rev. D} {\bf 95}, no. 8, 084004 (2017),
\href{https://arxiv.org/abs/1611.00590}{arXiv: 1611.00590}

\bibitem{Sheykhi:2008rk} 
A.~Sheykhi,
``Charged rotating dilaton black strings in AdS spaces,''
{\it Phys. Rev. D} {\bf 78}, 064055 (2008),
\href{\em https://arxiv.org/abs/0809.1130}{arXiv: 0809.1130}

\bibitem{horowitz1999} 
G.~T.~Horowitz and V.~E.~Hubeny,
``Quasinormal modes of AdS black holes and the approach to thermal equilibrium,''
{\it Phys. Rev. D} {\bf 62}, 024027 (2000),
\href{https://arxiv.org/abs/hep-th/9909056}{arXiv: hep-th/9909056}

\bibitem{adsQNM1997}
J.S.F. Chan, Robert B. Mann,,
``Scalar wave falloff in asymptotically anti-de Sitter backgrounds'',
{\em Phys.Rev.} {\bf D55} (1997) 7546-7562,
\href{https://arxiv.org/abs/gr-qc/9612026}{arXiv: gr-qc/9612026}.

\bibitem{adsQNM1999}
J.S.F. Chan, Robert B. Mann,
``Scalar wave falloff in topological black hole backgrounds'',
{\em Phys.Rev.} {\bf D59} (1999) 064025,
\href{http://dx.doi.org/10.1103/PhysRevD.59.064025}{DOI:10.1103/PhysRevD.59.064025}

\bibitem{konop0809} 
R.~A.~Konoplya and A.~Zhidenko,
``Stability of higher dimensional Reissner-Nordstrom-anti-de Sitter black holes,''
{\it Phys. Rev. D} {\bf 78}, 104017 (2008),
\href{https://arxiv.org/abs/0809.2048}{arXiv: 0809.2048}

\bibitem{dfzeng0605} 
D.~f.~Zeng, W.~s.~Xu and Y.~h.~Gao,
``Yang-type monopoles in 5 dimensional curved space-time,''
\href{https://arxiv.org/abs/hep-th/0605077}{arXiv: hep-th/0605077}

\end{thebibliography}
\end{document}